\documentstyle[12pt]{article}
\renewcommand{\theequation}{\arabic{section}.\arabic{equation}}
\newcommand{\eqreset}{\setcounter{equation}{0}}
\newtheorem{theorem}{Theorem}[section] 

\newtheorem{lemma}[theorem]{Lemma}

\begin{document}
\title{Analytic Bethe ansatz related to the  
 Lie superalgebra $C(s)$}
\author{Zengo Tsuboi
\\ 
Institute of Physics,                                   
 University of Tokyo,\\ Komaba 
  3-8-1, Meguro-ku, Tokyo 153-8902, Japan}
\date{}
\maketitle
\begin{abstract}
 An analytic Bethe ansatz is carried out related to 
 the type 1 Lie superalgebra $C(s)$. 
 We present eigenvalue formulae of transfer matrices 
 in dressed vacuum form (DVF) labeled by Young superdiagrams 
 with one row or one column. 
 We also propose an DVF related to
   a one parameter family of 
 finite dimensional irreducible representations.  
A class of transfer matrix functional 
 relations among these formulae is discussed. 
\end{abstract}
PACS numbers: 02.20.Qs, 02.20.Sv, 03.20.+i, 05.50.+q \\ 
Keywords:  Analytic Bethe Ansatz, Lie superalgebra, 
           Dressed vacuum form, 
           Solvable lattice model, Transfer matrix,  
           T-system. \\
      \\ 
\noindent Journal-ref: Physica A 267 (1999) 173-208 \\ 
DOI: 10.1016/S0378-4371(98)00673-6        

\ \\ 

\newpage
\eqreset
\section{Introduction}
 The Analytic Bethe ansatz \cite{R1,R2} is a powerful technique. 
We can postulates the eigenvalues of transfer matrices 
 in solvable lattice models associated with complicated representations  
 of underlying algebras, which are too hard to obtain by other method. 
In fact they are realized systematically in the dressed vacuum form (DVF)
  in terms of   
 Yangians $Y({\cal G})$ \cite{D} analogue of skew-Young tableaux 
 for ${\cal G}=A_{r}$, $B_{r}$, $C_{r}$, $D_{r}$ and 
 $G_{2}$ \cite{BR,KS1,KOS,S2}. 
 
Recently a similar analysis has been done \cite{T1,T2,T3} 
for the type 1 Lie superalgebra ${\cal G}=sl(r+1|s+1)$ \cite{Ka} case. 
 A set of DVFs are constructed and shown to satisfy a class of functional
  relations.
  
In this paper we will extend 
similar analyses to another type 1 Lie superalgebra 
${\cal G}=C(s)=osp(2|2s-2)$  case
\footnote{In the main text, we assume $s \ge 3$ 
($s \in {\bf Z}$). However, many formulae in the main text are 
formally valid for $C(2)$ case (see Appendix B). 
We note that 
solvable lattice models related to $C(2) \simeq sl(1|2)$ were 
studied in many literatures; see for example 
\cite{DFI,Ma,RM2,PF,BS,EK,FK,RM,T3}. 
}.
Solvable lattice models related to Lie superalgebras 
attract much interest in strongly correlated electron systems. 
The analytic Bethe ansatz will be powerful technique
 to analyze such systems. 
 
 Remarkable enough, we can express \cite{RW,OW,KOS} the 
 Bethe ansatz equation (BAE) 
 by the representation theoretical data of a Lie 
 algebra. 
  This is also the case with the Lie superalgebras: 
\cite{Kul} for $sl(r+1|s+1)$; \cite{RM} for 
$B(r|1)$, $B(0|s)$, $C(s)$ and $D(r|1)$. 
We assume, as our starting point, 
the BAE (\ref{BAE}) associated with the 
distinguished simple root system of $C(s)$ \cite{Ka}. 
 Then we execute an analytic Bethe ansatz  
axiomatically and construct 
 various kind of DVFs systematically. 
Our guiding principals to construct DVFs are  
{\em pole-freeness under the BAE} and {\em top term hypothesis} 
 \cite{KS1,KOS}.
 
 We define the functions ${\cal T}^{a}(u)$ (\ref{tate}) and 
 $T_{m}^{(1)}(u)$ (\ref{yoko}) as summations over certain tableaux 
labelled by Young superdiagrams 
with one column and one row respectively.
 They will be eigenvalue formulae of transfer 
matrices in DVFs generated by certain top terms. 
These top terms carry highest weights of some irreducible 
 representations of $C(s)$. 
 The function ${\cal T}^{1}(u)=T_{1}^{(1)}(u)$, 
 the simplest example of ${\cal T}^{a}(u)$ and 
 $T_{m}^{(1)}(u)$, corresponds to the eigenvalue formula of 
  a $C(s)$ vertex model in Ref. \cite{RM}. 
  We prove  pole-freeness of ${\cal T}^{a}(u)$ under the 
 BAE (\ref{BAE}), an essential property in the analytic Bethe ansatz. 
 This is based on the 
 assumption that the BAE (\ref{BAE})
is common to all the DVFs for transfer matrices with various
  fusion types in the auxiliary space 
  as long as they act on a common quantum space. 

The type 1 Lie superalgebra admits
 one parameter families of finite dimensional representations, 
which are not tensor-like. In the previous paper \cite{T3}, 
we gave the DVF related to a one parameter family of finite dimensional 
representations of $sl(r+1|s+1)$. On constructing this DVF, we 
deformed some DVF at which the top term carries 
the highest weight of a typical representation. 
 Based on the similar idea, we consider a deformation of the 
function $T_{s-1}^{(1)}(u)$ and construct a DVF with a
 continuous parameter. 
This deformation is compatible with the 
 top term hypothesis \cite{KS1,KOS}. 
We prove the pole freeness of this function. 
   We further define DVFs $T_{1}^{(a)}(u)$ ($a\in \{2,3,\dots, s\}$), 
   whose top term will carry \symbol{96}fundamental weights' 
   $\omega_{a}$.  
 We present a class of transfer matrix functional relations 
among DVFs. 
In particular, we conjecture a set of functional relations for 
 DVFs $T_{m}^{(a)}(u)$, which have $T_{-1}^{(1)}(u)$ and 
 $T_{1}^{(a)}(u)$ ($a\in\{2,3,\dots,s \}$) as 
 initial conditions. 
It may be viewed as a kind of the $T$-system \cite{KNS1} 
(see also \cite{BR,KS1,KOS,S2,T1,T2,T3,Ma,PF,KLWZ,K,KP,KS2,KNS2,KNH,
TK,T4,JKS}). 

The outline of this paper is given as follows.
In section 2, we briefly review the Lie superalgebra 
$C(s)$.
In section 3, we execute an analytic Bethe ansatz 
based upon the BAE (\ref{BAE}) associated 
with distinguished simple root system. 
We prove pole-freeness of the function 
${\cal T}^{a}(u)$. 
In section 4, we present an extension of the DVF defined in the 
previous section. 
In section 5, we mention functional relations for DVFs. 
 Section 6 is devoted to summary and discussion. 
Appendix A gives an example of the DVF related to $C(3)$. 
In Appendix B, we briefly mention a special case ($sl(1|2)$ case) of 
our previous results 
\cite{T1,T2,T3} and point out relation to $C(2)$ case.  
In this paper, we adopt similar notation in \cite{KS1,T1,T2,T3}.
\eqreset
\section{The Lie superalgebra $C(s)$}
In this section, we briefly review the 
Lie superalgebra $C(s)$ (see for example \cite{Ka,Ka2,BB,FJ,Go,Je}). 
A Lie superalgebra  is a ${\bf Z}_2$ graded algebra 
with a product, whose homogeneous
elements satisfy the graded Jacobi identity.

There are several choices of simple root systems depending on 
 the choices of Borel subalgebras.
The simplest system of simple roots is the so called
 distinguished one.
For example,  
the distinguished simple root system
 $\{\alpha_1,\dots,\alpha_{s} \}$
 of $C(s)$ has the following form (see Figure \ref{dynkin})
\begin{figure}
    \setlength{\unitlength}{1pt}
    \begin{center}
    \begin{picture}(250,50) 
      \put(22.929,12.9289){\line(1,1){14.14214}}
      \put(22.929,27.07107){\line(1,-1){14.14214}}
      \put(30,20){\circle{20}}
      \put(40,20){\line(1,0){20}}
      \put(70,20){\circle{20}}
      \put(80,20){\line(1,0){20}}
      \put(110,20){\line(1,0){10}}
      \put(130,20){\line(1,0){10}}
      \put(150,20){\line(1,0){20}}
      \put(180,20){\circle{20}}
      \put(188.8,25){\line(1,0){32.4}}
      \put(188.8,15){\line(1,0){32.4}}
      \put(230,20){\circle{20}}
      \put(27,0){$\alpha_{1}$}
      \put(67,0){$\alpha_{2}$}
      \put(177,0){$\alpha_{s-1}$}
      \put(227,0){$\alpha_{s}$}
      \put(190,20){\line(1,1){14.14214}}
      \put(190,20){\line(1,-1){14.14214}}
  \end{picture}
  \end{center}
  \caption{Dynkin diagram for the Lie superalgebra 
  $C(s)$ ($s \ge 3$) corresponding to the distinguished simple 
  root system: white circle denotes even root; 
   grey (a cross) circle denotes odd root $\alpha$ with 
   $(\alpha|\alpha)=0$.}
  \label{dynkin}
\end{figure}
\begin{eqnarray}
   && \alpha_{1} = \epsilon-\delta_{1} \nonumber  \\ 
   && \alpha_{i} = \delta_{i-1}-\delta_{i} ,
    \quad i=2,3,\dots,s-1, \\ 
   && \alpha_{s} =2\delta_{s-1} \nonumber
 \end{eqnarray}
where  
 $\epsilon;\delta_{1},\dots,\delta_{s-1}$ 
are the basis of the dual space of the Cartan subalgebra with
 the bilinear form $(\ |\ )$ such that 
\footnote{We normalized the longest simple root as 
$|(\alpha_{s}|\alpha_{s})|=2$.}
\begin{equation}
 (\epsilon|\epsilon)=\frac{1}{2}, \quad 
 (\epsilon|\delta_{j})=(\delta_{i}|\epsilon)=0 , \quad 
 (\delta_{i}|\delta_{j})=-\frac{\delta_{i\, j}}{2}  
\end{equation}
 $\{\alpha_i \}_{i \ne 1}$ are even roots and $\alpha_{1}$ 
is an odd root with $(\alpha_{1} | \alpha_{1})=0$.

Any weight can be expressed in the following form:
\begin{equation}
  \Lambda=\Lambda_{1} \epsilon
         +\sum_{j=1}^{s-1} \bar{\Lambda}_{j} \delta_{j},
\qquad \Lambda_{1},\bar{\Lambda_{j}}\in {\bf C}.
\end{equation}
We can rewrite this as follows: 
\begin{equation}
 \Lambda=b_{1}\omega_{1}+b_{2}\omega_{2}+\cdots +b_{s}\omega_{s},
\end{equation}
where $\omega_{a} $ is  a
\symbol{96}fundamental weight' 
\begin{equation}
 \omega_{a}=\left\{
   \begin{array}{lll}
    \epsilon & {\rm if} & a=1 \\ 
    -\epsilon +\delta_{1}+\delta_{2}+\cdots + \delta_{a-1} &
      {\rm if} & a\in \{2,3,\dots,s\}
   \end{array}
  \right.
\end{equation}  
 and $ b_{j} $ is Kac-Dynkin label 
\footnote{Note that Kac-Dynkin label of $\omega_{a}$ is 
$b_{j}=\delta_{a j}$.} 
\begin{eqnarray}
b_{j}=
\left\{
   \begin{array}{lll}
    \Lambda_{1}+\bar{\Lambda}_{1} & {\rm if} & j=1 \\ 
    \bar{\Lambda}_{j-1}-\bar{\Lambda}_{j} & {\rm if} & 
       j \in \{2,3,\dots ,s-1 \} \\ 
    \bar{\Lambda}_{s-1} & {\rm if } & j=s 
   \end{array}
  \right. .  
\end{eqnarray}
An irreducible representation $V(\Lambda)$ 
with the highest weight $\Lambda$
 is finite dimensional \cite{Ka2} if and only if 
\begin{equation}
  b_{j} \in {\bf Z}_{\ge 0} \qquad {\rm for} \quad j \ne 1.
\end{equation}
Note that $b_{1}$ can take on {\em complex} values. 

$V(\Lambda)$ is said to be typical if and only if 
\begin{equation}
  (\Lambda+\rho|\alpha)\ne 0 \quad {\rm for \quad all }
   \quad \alpha \in \Delta_{1}^{+},
\end{equation}
where $\Delta_{1}^{+}=\{ \epsilon \pm \delta_{j} \}$
; $\rho $ is the graded half sum of positive roots: 
 \begin{equation}
 \rho=\sum_{i=1}^{s-1}(s-i)\delta_{i}
      -(s-1)\epsilon. 
 \end{equation} 
There is a large class of finite dimensional representations,  
which is not tensor-like . 
For example, a one parameter family of finite dimensional  
representations with the highest weight 
\begin{eqnarray}
&& \Lambda (c)=c \epsilon = c \omega_{1},  \label{cweight} 
\end{eqnarray}
 is typical if 
\begin{equation}
  c \ne  0,1,\dots, s-2;\quad s,s+1,\dots,2s-2 . \label{typical}
\end{equation} 
Note that the first Kac-Dynkin label of $\Lambda (c)$ takes  
non-integer value if the parameter $c$ is  
 non-integral. 
 
 The dimensionality of the typical representation of 
 $C(s)$ with the highest weight 
 $\Lambda $ is given 
 \cite{Ka2,Je} as follows 
 \begin{eqnarray}
  {\rm dim}V(\Lambda)&=&2^{2(s-1)}
   \prod_{1\le i \le s-1}
  \frac{\bar{\Lambda}_{i}+s-i}{s-i}\nonumber \\
  & \times & \prod_{1 \le i< j \le s-1}
  \frac{(\bar{\Lambda}_{i}-\bar{\Lambda}_{j}+j-i)
         (\bar{\Lambda}_{i}+\bar{\Lambda}_{j}+2s-i-j)}
         {(j-i)(2s-i-j)}.
  \label{dim}
 \end{eqnarray}
 As for the atypical finite dimensional representation, 
 the dimensionality is given \cite{Je} as follows: 
\begin{eqnarray}
  {\rm dim}V(\Lambda)&=& \frac{2^{2s-3}}{(s-1)!}
   \{\prod_{1\le i \le s-1}
  \frac{(2i)!}{(s-1-i)!(s-1+i)!} \}
    \nonumber \\ 
&& \times (\prod_{1\le i \le s-1;i\ne k} x_{i}) 
    \{\prod_{1 \le i< j \le s-1;i,j\ne k}
    (x_{i}-x_{j})(x_{i}+x_{j}) \}
      \nonumber \\ 
&& \times (-1)^{k-1}
   \sum_{j=0}^{2s-3} \sum_{l=0}^{j}
   (-1)^{l}2^{-j}
 \left(
   \begin{array}{c}
   j \\ 
   l
   \end{array}
 \right)
   (x_{k}-l) 
   \nonumber \\ 
   && \times 
   \prod_{1 \le i \le s-1;i\ne k}
   (x_{k}-x_{i}-l)(x_{k}+x_{i}-l), 
\end{eqnarray}
where $x_{i}=\bar{\Lambda}_{i}+s-i$ for atypical 
with respect to 
$\epsilon+\delta_{k}$; 
$x_{i}=\bar{\Lambda}_{i}+s-i$ ($i\ne k$) 
 and $x_{k}=\bar{\Lambda}_{k}+s-1-k$ for atypical 
 with respect to $\epsilon-\delta_{k}$. 
\eqreset
\section{Analytic Bethe ansatz}
Consider the following type of the Bethe
 ansatz equation (cf \cite{RW,OW,Kul,RM,KOS}).
\begin{eqnarray}
 - \prod_{j=1}^{N}
 \frac{\Phi (u_k^{(a)}-w_{j}^{(a)}+\frac{b_{j}^{(a)}}{t_{a}})}
    {\Phi(u_k^{(a)}-w_{j}^{(a)}-\frac{b_{j}^{(a)}}{t_{a}})}
   &=&(-1)^{{\rm deg}(\alpha_a)} 
    \prod_{b=1}^{s}\frac{Q_{b}(u_k^{(a)}+(\alpha_a|\alpha_b))}
           {Q_{b}(u_k^{(a)}-(\alpha_a|\alpha_b))}, \label{BAE} \\ 
        Q_{a}(u)&=& \prod_{j=1}^{N_{a}}\Phi(u-u_j^{(a)}), 
        \label{Q_a} 
\end{eqnarray}
where 
$N,N_{a} \in {\bf Z }_{\ge 0}$; 
$u_{j}^{(a)}, w_{j}^{(a)}\in {\bf C}$; $a,k \in {\bf Z}$ 
($1\le a \le s$,$\ 1\le k\le N_{a}$);
 $t_{1}=2$; $t_{a}=-2 (2\le a \le s-1)$; 
 $t_{s}=-1 $;
  $b_{j}^{(a)} \in {\bf Z}_{\ge 0} (2\le a \le s)$;
   $b_{j}^{(1)} \in {\bf C}$  
  and  
\begin{eqnarray}
    {\rm deg}(\alpha_a)&=&\left\{
              \begin{array}{@{\,}ll}
                0  & \mbox{for even root} \\ 
                1 & \mbox{for odd root} 
              \end{array}
              \right. \\ 
             &=& \delta_{a,1}. \nonumber 
\end{eqnarray}
Here  $\Phi$ is a function, which has zero at $u=0$. 
For example, we have 
\begin{equation}
\Phi(u)=\frac{q^u-q^{-u}}{q-q^{-1}},
\end{equation}
where $q$ is generic.
The parameters in the left hand side of the BAE (\ref{BAE}) 
are the Kac-Dynkin labels 
\begin{equation}
(b_{j}^{(1)},b_{j}^{(2)},\dots,b_{j}^{(s)}) 
 \label{kac-dynkin-bae} 
\end{equation} 
of highest weight representations of $C(s)$, which 
are related to the quantum space. 

We define the sets 
\begin{eqnarray}
    &&  J=\{ 1,2,\dots,s,\bar{s},\dots,\bar{2},\bar{1}\},
    \nonumber \\
    &&  J_{+}=\{ 1,\bar{1}\}, \quad 
        J_{-}=\{2,3,\dots,s,\bar{s},\dots,\bar{3},\bar{2} \}   
  \label{set}
\end{eqnarray}
with the total order 
\begin{eqnarray} 
 1\prec 2 \prec \cdots \prec s 
 \prec \bar{s} \prec \cdots \prec \bar{2} 
 \prec \bar{1} \label{order}
\end{eqnarray}
and with the grading 
\begin{equation}
      p(a)=\left\{
              \begin{array}{@{\,}ll}
                0  & \mbox{for $a \in J_{+}$},  \\ 
                1 & \mbox{for $a \in J_{-}$ }.
              \end{array}
            \right. \label{grading}
\end{equation}
For $a \in J $, we define
\footnote{
In this paper, we often abbreviate the spectral parameter $u$.}
 the function 
\begin{eqnarray}
\hspace{-40pt} && \framebox{$1$}_{u}=\psi_{1}(u)
     \frac{Q_{1}(u-\frac{1}{2})}{Q_{1}(u+\frac{1}{2})} 
      ,  \nonumber \\ 
\hspace{-40pt} && \framebox{$a$}_{u}=\psi_{a}(u)
   \frac{Q_{a-1}(u-\frac{a-1}{2})Q_{a}(u-\frac{a-4}{2})}
        {Q_{a-1}(u-\frac{a-3}{2})Q_{a}(u-\frac{a-2}{2})}
      \qquad  2 \le a \le s-1,\nonumber \\
\hspace{-40pt} && \framebox{$s$}_{u}=\psi_{s}(u)
   \frac{Q_{s-1}(u-\frac{s-1}{2})Q_{s}(u-\frac{s-5}{2})}
        {Q_{s-1}(u-\frac{s-3}{2})Q_{s}(u-\frac{s-1}{2})},
        \nonumber \\
\hspace{-40pt} && \framebox{$\bar{s}$}_{u}=\psi_{\bar{s}}(u)
   \frac{Q_{s-1}(u-\frac{s-1}{2})Q_{s}(u-\frac{s+3}{2})}
        {Q_{s-1}(u-\frac{s+1}{2})Q_{s}(u-\frac{s-1}{2})},
        \nonumber \\
\hspace{-40pt} && \framebox{$\bar{a}$}_{u}=\psi_{\bar{a}}(u)
   \frac{Q_{a-1}(u-\frac{2s-a-1}{2})Q_{a}(u-\frac{2s-a+2}{2})}
        {Q_{a-1}(u-\frac{2s-a+1}{2})Q_{a}(u-\frac{2s-a}{2})}
      \qquad  2 \le a \le s-1,
      \nonumber \\
\hspace{-40pt} && \framebox{$\bar{1}$}_{u}=\psi_{\bar{1}}(u)
     \frac{Q_{1}(u-\frac{2s-3}{2})}{Q_{1}(u-\frac{2s-1}{2})} ,
   \label{z+}
\end{eqnarray}
where $Q_{0}(u)=Q_{s+1}(u)=1$. 
 In this section, we will consider the case,
  as an example, where  
  Kac-Dynkin labels in (\ref{kac-dynkin-bae}) have the form   
 $b_{j}^{(a)}=\delta_{a \ 1}$ ($1\le a\le s$).  
 In this case, the vacuum part of the function 
$\framebox{a}_{u}$ (\ref{z+}) takes the following form: 
\begin{eqnarray}
 \psi_{1}(u)&=&\phi(u+1)\phi(u-s+1), \nonumber \\ 
 \psi_{a}(u)&=&\phi(u)\phi(u-s+1) 
   \qquad 2 \preceq a \preceq \bar{2}, \nonumber \\
 \psi_{\bar{1}}(u)&=&\phi(u)\phi(u-s), 
   \label{psi}
\end{eqnarray}
where 
\begin{eqnarray}
 \phi(u)=\prod_{j=1}^{N}\Phi(u-w_{j}^{(1)}).
\end{eqnarray}
The generalization to the case of arbitrary Kac-Dynkin labels 
 (\ref{kac-dynkin-bae}) will 
be achieved by suitable redefinition of the function $\psi_{a}(u)$, and  
such redefinition will not essentially   
 influence the subsequent argument.  
  
 Under the BAE (\ref{BAE}),
 we have
 \footnote{
 Here $Res_{u=a}f(u)$ denotes the residue of a function $f(u)$ at $u=a $.
 } 
\begin{eqnarray}
\hspace{-45pt}&& Res_{u=-\frac{1}{2}+u_{k}^{(1)}}
 (\framebox{$1$}_{u}-\framebox{$2$}_{u})=0,
  \label{res1} \\ 
\hspace{-45pt}&& Res_{u=\frac{d-2}{2}+u_{k}^{(d)}}
 (\framebox{$d$}_{u}+\framebox{$d+1$}_{u})=0 ,
    \qquad 2\le d \le s-1 , \\
\hspace{-45pt}&& Res_{u=\frac{s-1}{2}+u_{k}^{(s)}}
(\framebox{$s$}_{u}+\framebox{$\bar{s}$}_{u})=0  ,
      \\
\hspace{-45pt}&& Res_{u=\frac{2s-d}{2}+u_{k}^{(d)}}
   (\framebox{$\bar{d}$}_{u}+\framebox{$\overline{d+1}$}_{u})=0,
    \qquad 2\le d \le s-1  ,\\
\hspace{-45pt}&& Res_{u=\frac{2s-1}{2}+u_{k}^{(1)}}
                 (\framebox{$\bar{1}$}_{u}-\framebox{$\bar{2}$}_{u})=0.
                 \label{res5} 
\end{eqnarray}
%
We shall present functions with spectral parameter $u\in {\rm C}$, 
 which are candidates of DVFs of various fusion types in the 
 auxiliary space of transfer matrices of 
 $U_{q}(C(s)^{(1)})$ vertex model. 
For $a \in {\bf Z}_{\ge 0}$, 
we define the DVF labelled by the Young superdiagram 
\footnote{Note that 
Kac-Dynkin label $(b_{1},b_{2},\dots ,  b_{s})$ is related 
  to 
the Young superdiagram with shape $(\mu_{1},\mu_{2},\dots)$ 
as follows 
\begin{eqnarray}
  b_{1} &=& \mu_{1} +\eta_{1}, \nonumber \\ 
  b_{1+i} &=& \eta_{i} -\eta_{i+1} \qquad 
  {\rm for} \quad i\in \{1,2,\dots,s-2\}, \\ 
  b_{s} &=& \eta_{s-1} \nonumber,
\end{eqnarray} 
  where $\eta_{i}=Max\{\mu_{i}^{\prime}-1,0 \} $. 
}
with shape $(1^{a})$ as a summation
\footnote{We assume ${\cal T}^{0}(u)=1. $}
 over products of 
the boxes in (\ref{z+}) : 
\begin{eqnarray}
{\cal T}^{a}(u) &=&
 \sum_{\{i_{k}\}\in B(1^{a})}
  (-1)^{\sum_{j=1}^{a} p(i_{j})}
  \begin{array}{|c|}\hline 
     	i_{1}  \\ \hline
      i_{2}  \\ \hline
     	\vdots \\ \hline 
     	i_{a}  \\ \hline
  \end{array},
  \label{tate} 
\end{eqnarray}
where the spectral parameter $u$ is shifted as 
$u+\frac{a-1}{2},u+\frac{a-3}{2},\dots ,u-\frac{a-1}{2}$ 
from top to the bottom. 
$B(1^{a})$ is a set of tableaux $\{i_{k}\}$ 
obeying the following rule 
(admissibility conditions) 
\begin{enumerate}
\item 
For $i_{k},i_{k+1} \in J_{+}$, 
\begin{equation}
 i_{k} \prec i_{k+1} \label{adtate1}
\end{equation}
\item 
and for  $i_{k},i_{k+1} \in J$,
\begin{equation}   
i_{k} \preceq i_{k+1}
\end{equation}
unless 
\begin{equation}
 i_{k}=\bar{s}, \qquad i_{k+1}=s .\label{adtate3}
\end{equation}
\end{enumerate}
The top term
\footnote{Here we omit the vacuum part.} 
\cite{KS1} of the DVF (\ref{tate}) will be 
\begin{eqnarray}
 \left.
  (-1)^{a-1}
  \begin{array}{|c|}\hline 
     	1  \\ \hline
        2  \\ \hline
     	\vdots \\ \hline 
     	2  \\ \hline
  \end{array}
  \, 
 \right\}
 \! 
 {\tiny a} \; 
  =(-1)^{a-1}
\frac{Q_{1}(u-\frac{a}{2})Q_{2}(u+\frac{a-1}{2})}
     {Q_{1}(u+\frac{a}{2})Q_{2}(u-\frac{a-1}{2})},
\end{eqnarray}
which carries $C(s)$ weight $\epsilon + (a-1)\delta_{1}
=a\omega_{1}+(a-1)\omega_{2}$. 
We believe the DVF (\ref{tate}) is generated from this term. 

\begin{table}
\begin{center}
 \begin{tabular}{|c|c|c|c|c|c|c|c|} \hline 
 $a$     & 0 & 1 & 2  & 3  & 4   & 5   & 6   \\ \hline 
$\sharp B(1^a)$ & 1 & 6 & 20 & 50 & 105 & 196 & 336 \\ \hline 
 \end{tabular} 
\end{center}
\caption{The number of the terms 
 in ${\cal T}^{a}(u)$ 
 for $C(3)$.}
\label{num-ta} 
\end{table}
The number of the terms in ${\cal T}^{a}(u)$ is given as 
  (see Table \ref{num-ta}) 
\begin{eqnarray}
 \sharp B(1^a)= \sum_{n=0}^{[\frac{a}{2}]}
  {\cal D}(a,n),
\end{eqnarray}
where 
\begin{eqnarray}
  {\cal D}(a,n)&=&
  \sum_{k=0}^{a-2n}
  \left\{
  \left(
   \begin{array}{c}
     k+s-2 \\ 
     k
   \end{array}
    \right)
   +
   \left(
   \begin{array}{c}
     k+s-3 \\ 
     k-1
   \end{array}
   \right)
  \right\} 
  \nonumber \\ 
  && \hspace{-30pt} \times 
  \left\{
  \left(
   \begin{array}{c}
     a-2n-k+s-2 \\ 
     a-2n-k
   \end{array}
    \right)
   +
   \left(
   \begin{array}{c}
     a-2n-k+s-3 \\ 
     a-2n-k-1
   \end{array}
   \right)
  \right\}.
\end{eqnarray} 
\begin{table}
\begin{center}
 \begin{tabular}{|c|c||c|c||c|c||c|c||c|c|} \hline 
 $\{ b_{j} \}$ & dim. & 
 $\{ b_{j} \}$ & dim. & 
 $\{ b_{j} \}$ & dim. & 
 $\{ b_{j} \}$ & dim. &
 $\{ b_{j} \}$ & dim.    \\ \hline
 0 0 0 & 1  & $k$ 0 0 & 16  & 0 5 0 & 351 & 0 0 5 & 286 & 6 5 0 & 231 
  \\ \hline 
 1 0 0 & 6  & 0 1 0 & 15  & 0 0 1 & 10  & 2 1 0 & 19 & -2 1 0 & 64  
  \\ \hline 
 2 0 0 & 16 & 0 2 0 & 49  & 0 0 2 & 35  & 3 2 0 & 44 & -4 1 0 & 64 
  \\ \hline 
 3 0 0 & 10 & 0 3 0 & 111 & 0 0 3 & 84  & 4 3 0 & 85 & -2 2 0 & 160 
  \\ \hline 
 4 0 0 & 15 & 0 4 0 & 209 & 0 0 4 & 165 & 5 4 0 & 146& -2 3 0 & 320 
  \\ \hline 
  \end{tabular} 
\end{center}
\caption{Dimensionality of the module  
  $V(b_{1}\omega_{1}+b_{2}\omega_{2}+b_{3}\omega_{3})$ 
 for $C(3)$. Here $k \ne 0,1,3,4 ; k \in {\bf R}$; 
 $\{b_{j}\}$ is Kac-Dynkin label $\{b_{1},b_{2},b_{3}\}$.}
\label{dim-c3} 
\end{table}
We have checked the following relation for several cases
 (see Table \ref{dim-c3}), 
\begin{equation}
{\cal D}(a,n)=
\left\{
\begin{array}{lll}
  {\rm dim} V(\epsilon + (a-2n-1)\delta_{1}) 
  & {\rm if } & n \in \{0,1,\dots, [\frac{a}{2}]-1 \}, \\ 
  {\rm dim}V(0) & {\rm if }& n=\frac{a}{2} , a \in 2{\bf Z}_{\ge 0}, \\ 
  {\rm dim}V(\epsilon) & {\rm if }& n=\frac{a-1}{2} ,
     a \in 2{\bf Z}_{\ge 0}+1.
\end{array}
\right.  
\end{equation}
This relation  suggests a possibility 
that the auxiliary space  $W_{(1^{a})}$ 
\footnote{It will be a module of super Yangian or 
quantum affine superalgebra
 \cite{Y1,Y2,DGLZ2}.}
of the DVF ${\cal T}^{a}(u)$ 
decomposes as a $C(s)$ module as follows:
\begin{eqnarray}
 W_{(1^{a})} &\simeq&
  V(\epsilon + (a-1)\delta_{1}) \oplus 
  V(\epsilon + (a-3)\delta_{1}) \oplus
  \cdots 
  \nonumber \\ 
  && \hspace{80pt} \cdots \oplus 
    \left\{
      \begin{array}{lll}
        V(0) & {\rm if} & a \in 2{\bf Z}_{\ge 0}, \\ 
        V(\epsilon ) & {\rm if} & a \in 2{\bf Z}_{\ge 0}+1.
      \end{array}
    \right. 
    \label{decom}
\end{eqnarray}
This relation seem to suggest a 
superization of Kirillov-Reshetikhin formula 
\cite{KR}, which gives multiplicity of occurrence of the irreducible 
representations of Lie superalgebra in Yangian module.  

 For $m\in \{1,2,\dots, s-1 \}$, 
we also define the DVF labelled by the Young superdiagram 
with shape $(m^{1})$ as follows
\footnote{We assume $T_{0}^{(1)}(u)=1$.}
: 
\begin{eqnarray}
T_{m}^{(1)}(u) &=&
 \sum_{\{i_{k}\}\in B(m^{1})}
  (-1)^{\sum_{j=1}^{m} p(i_{j})}
  \begin{array}{|c|c|c|c|}\hline 
     	i_{1} & i_{2} & \cdots & i_{m} \\ \hline
  \end{array}
  \label{yoko} ,
\end{eqnarray}
where the spectral parameter $u$ is shifted as 
$u-\frac{m-1}{2},u-\frac{m-3}{2},\dots ,u+\frac{m-1}{2}$
from left to right. 
$B(m^{1})$  is a set of tableaux $\{i_{k}\}$ 
obeying the following rule 
(admissibility conditions): 
\begin{enumerate}
\item 
 For $i_{k},i_{k+1} \in J_{-}$, 
\begin{equation}
 i_{k} \prec i_{k+1} \label{adm1} 
\end{equation}
and 
\begin{equation}
 s+j-k \ge d 
\quad 
{\rm provided} 
\quad
 i_{j}=d \quad 
 {\rm and} \quad 
 i_{k}=\bar{d},
\end{equation}
\item 
for $i_{k},i_{k+1} \in J$
\begin{equation}   
i_{k} \preceq i_{k+1}. \label{adm3} 
\end{equation}
\end{enumerate}
The top term 
\footnote{Here we omit the vacuum part.} 
 \cite{KS1} of the DVF (\ref{yoko}) will be 
\begin{eqnarray}
\underbrace{
 \begin{array}{|c|c|c|c|}\hline 
     	1 & 1 & \cdots & 1 \\ \hline
  \end{array}
 }_{m}
=\frac{Q_{1}(u-\frac{m}{2})}{Q_{1}(u+\frac{m}{2})},
\end{eqnarray}
which carries $C(s)$ weight $m\epsilon =m\omega_{1}$.  
We believe the DVF (\ref{yoko}) is generated from this term. 

The number ${\cal N}_{m}^{(1)}$ 
of the terms in $T_{m}^{(1)}(u)$ 
for $m\in \{0,1,2,\dots,s-1 \}$
\footnote{As for $m\ge s (m\in {\bf Z})$, see section 4.}
is given as
 (see Table \ref{numt1-3}) 
\begin{eqnarray}
 \sharp B(m^1) &=&
  \sum_{k=0}^{m}
  \left\{
  \left(
   \begin{array}{c}
     2s-2 \\ 
     k
   \end{array}
    \right)
   -
   \left(
   \begin{array}{c}
     2s-2 \\ 
     k-2
   \end{array}
   \right)
  \right\} 
 (m-k+1)
\end{eqnarray} 
\begin{table}
\begin{center}
 \begin{tabular}{|c|c|c|c|c|c|c|} \hline 
 $m$
                   & $0$ & $1$ & $2$  & $3$ & $4$ & otherwise \\ \hline 
${\cal N}_{m}^{(1)}$ & $1$ & $6$ & $16$ & $10$  & $15$  & $16$ \\ \hline 
  \end{tabular} 
\end{center}
\caption{The number of the terms 
 in $T_{m}^{(1)}(u)$ 
 for $C(3)$.}
\label{numt1-3} 
\end{table}
We have checked the relation $ \sharp B(m^1)={\rm dim} V(m\omega_{1})$ 
for several cases. 
This relation  suggests a possibility that 
the auxiliary space  $W_{(m^{1})}$ of the DVF $ T_{m}^{(1)}(u)$ 
is $W_{(m^{1})} \simeq V(m\epsilon)$ as a $C(s)$ module. 

Note that the function ${\cal T}^{1}(u)=T_{1}^{(1)}(u)$ 
coincides with  
the eigenvalue formula by the algebraic Bethe ansatz
 \cite{RM}. 

We remark that $\framebox{$a$}_{u}$ is transformed to 
$ \framebox{$\overline{a}$}_{u}$ 
under the following transformation:  
\begin{eqnarray}
&& u \to -(u-s+1), \nonumber \\ 
&& u_{j}^{(a)} \to -u_{j}^{(a)}, \label{cross} \\ 
&& w_{j}^{(1)} \to -w_{j}^{(1)}, \nonumber 
\end{eqnarray}
if $\Phi (-u) = \pm \Phi (u)$. In this case, 
${\cal T}^{a}(u)$ (\ref{tate}) and $T_{m}^{(1)}(u)$ (\ref{yoko}) 
are invariant under the transformation (\ref{cross}). 

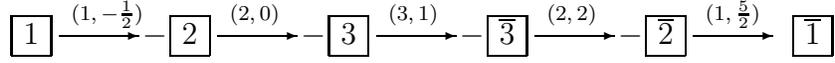
\begin{figure}
    \setlength{\unitlength}{1.5pt}
    \begin{center}
    \begin{picture}(207,40) 
      \put(0,0){\line(1,0){10}}
      \put(10,0){\line(0,1){10}}
      \put(10,10){\line(-1,0){10}}
      \put(0,10){\line(0,-1){10}}
      \put(3,3){$1$}
      \put(12,5){\vector(1,0){20}}
      \put(15,9){\scriptsize{$(1,-\frac{1}{2})$}}
      \put(33,3){$-$}
      \put(40,0){\line(1,0){10}}
      \put(50,0){\line(0,1){10}}
      \put(50,10){\line(-1,0){10}}
      \put(40,10){\line(0,-1){10}}
      \put(43,3){$2$}
      \put(52,5){\vector(1,0){20}}
      \put(55,9){\scriptsize{$(2,0)$}}
      \put(73,3){$-$}
      \put(80,0){\line(1,0){10}}
      \put(90,0){\line(0,1){10}}
      \put(90,10){\line(-1,0){10}}
      \put(80,10){\line(0,-1){10}}
      \put(83,3){$3$}
      \put(92,5){\vector(1,0){20}}
      \put(95,9){\scriptsize{$(3,1)$}}
      \put(113,3){$-$}
      \put(120,0){\line(1,0){10}}
      \put(130,0){\line(0,1){10}}
      \put(130,10){\line(-1,0){10}}
      \put(120,10){\line(0,-1){10}}
      \put(123,3){$\overline{3}$}
      \put(132,5){\vector(1,0){20}}
      \put(135,9){\scriptsize{$(2,2)$}}
      \put(153,3){$-$}
      \put(160,0){\line(1,0){10}}
      \put(170,0){\line(0,1){10}}
      \put(170,10){\line(-1,0){10}}
      \put(160,10){\line(0,-1){10}}
      \put(163,3){$\overline{2}$}
      \put(172,5){\vector(1,0){20}}
      \put(175,9){\scriptsize{$(1,\frac{5}{2})$}}
      \put(197,0){\line(1,0){10}}
      \put(207,0){\line(0,1){10}}
      \put(207,10){\line(-1,0){10}}
      \put(197,10){\line(0,-1){10}}
      \put(200,3){$\overline{1}$}
  \end{picture}
  \end{center}
  \caption{The  Bethe-strap structure of the 
  DVF $T_{1}^{(1)}(u)$  (\ref{t1-ex}) for 
   the Lie superalgebra $C(3)$:  
 The pair $(a,b)$ denotes the common pole $u_{k}^{(a)}+b$ of the pair   
 of the tableaux connected by the arrow.   
 This common pole vanishes under the BAE (\ref{BAE2}).
 The leftmost (resp. rightmost) tableau corresponds to the 
 \symbol{96}highest weight vector\symbol{39}
 (resp.\symbol{96}lowest weight vector\symbol{39}), 
 which is called the {\em top term} (resp. {\em bottom term }). 
 Such a  correspondence between certain term in the DVF and a highest 
 weight (to be more precise, a kind of Drinfel'd polynomial)
  may be called {\em top term hypothesis} \cite{KS1,KOS}.}
  \label{best1}
\end{figure}

Now we shall present examples of (\ref{tate}) and (\ref{yoko}) 
for  $C(3), J_{+}=\{1,\overline{1} \},
J_{-}=\{2,3,\overline{3},\overline{2} \}$ case: 
\begin{eqnarray}
{\cal T}^{1}(u) &=&T_{1}^{(1)}(u)=
  \begin{array}{|c|}\hline 
     	1  \\ \hline
  \end{array}
 -\begin{array}{|c|}\hline 
  	2  \\ \hline
  \end{array}
 -\begin{array}{|c|}\hline 
  	3  \\ \hline
  \end{array}
 -\begin{array}{|c|}\hline 
  	\overline{3}  \\ \hline
  \end{array}
 -\begin{array}{|c|}\hline 
  	\overline{2}  \\ \hline
  \end{array}
 +\begin{array}{|c|}\hline 
  	\overline{1}  \\ \hline
  \end{array}
\nonumber \\ 
&=&
\phi(-2 + u)\phi(1 + u)
\frac{Q_{1}(-\frac{1}{2} + u)}{Q_{1}(\frac{1}{2} + u)}
 \nonumber \\ 
&-&\phi(-2 + u)\phi(u)\frac{Q_{1}(-\frac{1}{2} + u)Q_{2}(1 + u)}
    {Q_{1}(\frac{1}{2} + u)Q_{2}(u)}
 \nonumber \\ 
&-& \phi(-2 + u)\phi(u)\frac{Q_{2}(-1 + u)Q_{3}(1 + u)}
   {Q_{2}(u)Q_{3}(-1 + u)}
   \nonumber \\
&-&\phi(-2 + u)\phi(u)\frac{Q_{2}(-1 + u)Q_{3}(-3 + u)}
  {Q_{2}(-2 + u)Q_{3}(-1 + u)}
  \nonumber \\ 
&-&\phi(-2 + u)\phi(u)\frac{Q_{1}(-\frac{3}{2} + u)Q_{2}(-3 + u)}
     {Q_{1}(-{\frac{5}{2}} + u)Q_{2}(-2 + u)}
     \nonumber \\ 
&+& \phi(-3 + u)\phi(u)\frac{Q_{1}(-\frac{3}{2} + u)}
  {Q_{1}(-\frac{5}{2} + u)}, 
\label{t1-ex}
\end{eqnarray}
\begin{eqnarray}
{\cal T}^{2}(u) &=&
  -\begin{array}{|c|}\hline 
     	1  \\ \hline
     	2  \\ \hline 
   \end{array}
  -\begin{array}{|c|}\hline 
     	1  \\ \hline
     	3  \\ \hline 
   \end{array}
  -\begin{array}{|c|}\hline 
     	1  \\ \hline
     	\overline{3}  \\ \hline 
   \end{array}
  -\begin{array}{|c|}\hline 
     	1  \\ \hline
     	\overline{2}  \\ \hline 
   \end{array}
  +\begin{array}{|c|}\hline 
     	1  \\ \hline
     	\overline{1}  \\ \hline 
   \end{array}
  +\begin{array}{|c|}\hline 
     	2  \\ \hline
     	2  \\ \hline 
   \end{array}
  +\begin{array}{|c|}\hline 
     	2  \\ \hline
     	3  \\ \hline 
   \end{array}
  +\begin{array}{|c|}\hline 
     	2  \\ \hline
     	\overline{3}  \\ \hline 
   \end{array}
  +\begin{array}{|c|}\hline 
     	2  \\ \hline
     	\overline{2}  \\ \hline 
   \end{array}
  -\begin{array}{|c|}\hline 
     	2  \\ \hline
     	\overline{1}  \\ \hline 
   \end{array}
   \nonumber \\ 
  &+&\begin{array}{|c|}\hline 
     	3  \\ \hline
     	3  \\ \hline 
   \end{array}
  +\begin{array}{|c|}\hline 
     	3  \\ \hline
     	\overline{3}  \\ \hline 
   \end{array}
  +\begin{array}{|c|}\hline 
     	3  \\ \hline
     	\overline{2}  \\ \hline 
   \end{array}
  -\begin{array}{|c|}\hline 
     	3  \\ \hline
     	\overline{1}  \\ \hline 
   \end{array}
  +\begin{array}{|c|}\hline 
     	\overline{3}  \\ \hline
     	\overline{3}  \\ \hline 
   \end{array}
  +\begin{array}{|c|}\hline 
     	\overline{3}  \\ \hline
     	\overline{2}  \\ \hline 
   \end{array}
  -\begin{array}{|c|}\hline 
     	\overline{3}  \\ \hline
     	\overline{1}  \\ \hline 
   \end{array}
  +\begin{array}{|c|}\hline 
     	\overline{2}  \\ \hline
     	\overline{2}  \\ \hline 
   \end{array}
  -\begin{array}{|c|}\hline 
     	\overline{2}  \\ \hline
     	\overline{1}  \\ \hline 
   \end{array}
   \nonumber \\ 
  &+&\begin{array}{|c|}\hline 
     	\overline{3}  \\ \hline
     	3  \\ \hline 
   \end{array}
   \nonumber \\
&=& \phi(-\frac{3}{2} + u)\phi(-\frac{1}{2} + u) 
\left\{ \right.\nonumber \\  
 &-& \phi(-\frac{5}{2} + u)\phi(\frac{3}{2} + u)
       \frac{Q_{1}(-1 + u)
       Q_{2}(\frac{1}{2} + u)}{Q_{1}(1 + u)Q_{2}(-\frac{1}{2} + u)}
       \nonumber \\ 
&-&\phi(-\frac{5}{2} + u)\phi(\frac{3}{2} + u)
       \frac{Q_{1}(u)
       Q_{2}(-\frac{3}{2} + u)Q_{3}(\frac{1}{2} + u)}{Q_{1}(1 + u)
       Q_{2}(-\frac{1}{2} + u)Q_{3}(-\frac{3}{2} + u)} 
       \nonumber \\ 
&-&\phi(-\frac{5}{2} + u)\phi(\frac{3}{2} + u)
       \frac{Q(1,u)
       Q_{2}(-\frac{3}{2} + u)Q_{3}(-\frac{7}{2} + u)}{Q_{1}(1 + u)
       Q_{2}(-\frac{5}{2} + u)Q_{3}(-\frac{3}{2} + u)}
       \nonumber \\ 
&-&\phi(-\frac{5}{2} + u)\phi(\frac{3}{2} + u)
  \frac{Q_{1}(-2 + u)
  Q_{1}(u)Q_{2}(-\frac{7}{2} + u)}
  {Q_{1}(-3 + u)Q_{1}(1 + u)Q_{2}(-\frac{5}{2} + u)}
      \nonumber \\ 
&+& \phi(-\frac{7}{2} + u)\phi(\frac{3}{2} + u)
  \frac{Q_{1}(-2 + u)Q_{1}(u)}
  {Q_{1}(-3 + u)Q_{1}(1 + u)}
  \nonumber \\ 
&+& \phi(-\frac{5}{2} + u)\phi(\frac{1}{2} + u)
  \frac{Q_{1}(-1 + u)
  Q_{2}(\frac{3}{2} + u)}
  {Q_{1}(1 + u)Q_{2}(-\frac{1}{2} + u)}
  \nonumber \\ 
&+& \phi(-\frac{5}{2} + u)\phi(\frac{1}{2} + u)
  \frac{Q_{1}(u)
      Q_{2}(-\frac{3}{2} + u)Q_{2}(\frac{3}{2} + u)
      Q_{3}(\frac{1}{2} + u)}
      {Q_{1}(1 + u)Q_{2}(-\frac{1}{2} + u)Q_{2}(\frac{1}{2} + u)
      Q_{3}(-\frac{3}{2} + u)}
      \nonumber \\ 
&+& \phi(-\frac{5}{2} + u)\phi(\frac{1}{2} + u)
      \frac{Q_{1}(u)Q_{2}(-\frac{3}{2} + u)Q_{2}(\frac{3}{2} + u)
      Q_{3}(-\frac{7}{2} + u)}{Q_{1}(1 + u)Q_{2}(-\frac{5}{2} + u)
      Q_{2}(\frac{1}{2} + u)Q_{3}(-\frac{3}{2} + u)}
      \nonumber \\ 
&+& \phi(-\frac{5}{2} + u)\phi(\frac{1}{2} + u)
  \frac{Q_{1}(-2 + u)Q_{1}(u)Q_{2}(-\frac{7}{2} + u)
  Q_{2}(\frac{3}{2} + u)}{Q_{1}(-3 + u)Q_{1}(1 + u)
      Q_{2}(-\frac{5}{2} + u)Q_{2}(\frac{1}{2} + u)}
      \nonumber \\ 
&-& \phi(-\frac{7}{2} + u)\phi(\frac{1}{2} + u)
   \frac{Q_{1}(-2 + u)Q_{1}(u)Q_{2}(\frac{3}{2} + u)}
  {Q_{1}(-3 + u)Q_{1}(1 + u)Q_{2}(\frac{1}{2} + u)}
  \nonumber \\ 
&+& \phi(-\frac{5}{2} + u)\phi(\frac{1}{2} + u)
   \frac{Q_{2}(-\frac{3}{2} + u)Q_{3}(\frac{1}{2} + u)
  Q_{3}(\frac{3}{2} + u)}{Q_{2}(\frac{1}{2} + u)
  Q_{3}(-\frac{3}{2} + u)Q_{3}(-\frac{1}{2} + u)}
  \nonumber \\ 
&+& \phi(-\frac{5}{2} + u)\phi(\frac{1}{2} + u)
   \frac{Q_{2}(-\frac{3}{2} + u)Q_{2}(-\frac{1}{2} + u)
   Q_{3}(-\frac{7}{2} + u)Q_{3}(\frac{3}{2} + u)}
  {Q_{2}(-\frac{5}{2} + u)Q_{2}(\frac{1}{2} + u)
  Q_{3}(-\frac{3}{2} + u)Q_{3}(-\frac{1}{2} + u)}
  \nonumber \\ 
&+& \phi(-\frac{5}{2} + u)\phi(\frac{1}{2} + u)
      \frac{Q_{1}(-2 + u)Q_{2}(-\frac{7}{2} + u)Q_{2}(-\frac{1}{2} + u)
      Q_{3}(\frac{3}{2} + u)}{Q_{1}(-3 + u)Q_{2}(-\frac{5}{2} + u)
      Q_{2}(\frac{1}{2} + u)Q_{3}(-\frac{1}{2} + u)}
      \nonumber \\ 
&-& \phi(-\frac{7}{2} + u)\phi(\frac{1}{2} + u)
  \frac{Q_{1}(-2 + u)
       Q_{2}(-\frac{1}{2} + u)Q_{3}(\frac{3}{2} + u)}{Q_{1}(-3 + u)
       Q_{2}(\frac{1}{2} + u)Q_{3}(-\frac{1}{2} + u)}
       \nonumber \\ 
&+& \phi(-\frac{5}{2} + u)\phi(\frac{1}{2} + u)
  \frac{Q_{2}(-\frac{1}{2} + u)Q_{3}(-\frac{7}{2} + u)
  Q_{3}(-\frac{5}{2} + u)}{Q_{2}(-\frac{5}{2} + u)
      Q_{3}(-\frac{3}{2} + u)Q_{3}(-\frac{1}{2} + u)}
      \nonumber \\ 
&+& \phi(-\frac{5}{2} + u)\phi(\frac{1}{2} + u)
   \frac{Q_{1}(-2 + u)
      Q_{2}(-\frac{7}{2} + u)Q_{2}(-\frac{1}{2} + u)
      Q_{3}(-\frac{5}{2} + u)}
      {Q_{1}(-3 + u)Q_{2}(-\frac{5}{2} + u)Q_{2}(-\frac{3}{2} + u)
      Q_{3}(-\frac{1}{2} + u)}
      \nonumber \\ 
&-& \phi(-\frac{7}{2} + u)
       \phi(\frac{1}{2} + u)
  \frac{Q_{1}(-2 + u)Q_{2}(-\frac{1}{2} + u)
       Q_{3}(-\frac{5}{2} + u)}{Q_{1}(-3 + u)Q_{2}(-\frac{3}{2} + u)
       Q_{3}(-\frac{1}{2} + u)}
       \nonumber \\ 
&+& \phi(-\frac{5}{2} + u)
      \phi(\frac{1}{2} + u)
  \frac{Q_{1}(-1 + u)Q_{2}(-\frac{7}{2} + u)}
      {Q_{1}(-3 + u)Q_{2}(-\frac{3}{2} + u)}
      \nonumber \\ 
 &-& \phi(-\frac{7}{2} + u)
       \phi(\frac{1}{2} + u)
  \frac{Q_{1}(-1 + u)
       Q_{2}(-\frac{5}{2} + u)}{Q_{1}(-3 + u)
       Q_{2}(-\frac{3}{2} + u)}
       \nonumber \\ 
 &+& \phi(-\frac{5}{2} + u)
      \phi(\frac{1}{2} + u)
  \frac{Q_{3}(-\frac{5}{2} + u)
      Q_{3}(\frac{1}{2} + u)}{
      Q_{3}(-\frac{3}{2} + u)Q_{3}(-\frac{1}{2} + u)}
     \left. \right\},
     \label{t2-ex}
\end{eqnarray}
\begin{eqnarray}
T_{2}^{(1)}(u) &=&
\begin{array}{|c|c|} \hline 
    1 & 1 \\ \hline 
\end{array}
-
\begin{array}{|c|c|}\hline
    1 & 2 \\ \hline 
\end{array}
-
\begin{array}{|c|c|}\hline
    1 & 3  \\ \hline
\end{array}
-
\begin{array}{|c|c|}\hline
    1 & \bar{3}  \\ \hline
\end{array}
-
\begin{array}{|c|c|}\hline
    1 & \bar{2}  \\ \hline
\end{array}
+
\begin{array}{|c|c|}\hline
    1 & \bar{1}  \\ \hline
\end{array} 
\nonumber \\ 
&+&
\begin{array}{|c|c|}\hline
    2 & 3  \\ \hline
\end{array}
+
\begin{array}{|c|c|}\hline
    2 & \bar{3}  \\ \hline
\end{array} 
+
\begin{array}{|c|c|}\hline
    2 & \bar{2}  \\ \hline
\end{array}
-
\begin{array}{|c|c|}\hline
    2 & \bar{1}  \\ \hline
\end{array}
+
\begin{array}{|c|c|}\hline
    3 & \bar{2}  \\ \hline
\end{array}
-
\begin{array}{|c|c|}\hline
    3 & \bar{1}  \\ \hline
\end{array}
\nonumber \\ 
&+&
\begin{array}{|c|c|}\hline
    \bar{3} & \bar{2}  \\ \hline
\end{array}
-
\begin{array}{|c|c|}\hline
    \bar{3} & \bar{1}  \\ \hline
\end{array}
-
\begin{array}{|c|c|}\hline
    \bar{2} & \bar{1}  \\ \hline
\end{array}
+
\begin{array}{|c|c|}\hline
    \bar{1} & \bar{1}  \\ \hline
\end{array}
\nonumber \\ 
&=& \phi(-\frac{5}{2} + u)\phi(\frac{1}{2} + u)
\{
 \phi(-\frac{3}{2} + u)\phi(\frac{3}{2} + u)
 \frac{Q_{1}(-1 + u)}{Q_{1}(1 + u)} 
 \nonumber \\ 
  &-& \phi(-\frac{3}{2}+ u)\phi(\frac{1}{2} + u)
   \frac{Q_{1}(-1 + u)Q_{2}(\frac{3}{2}+u)}
        {Q_{1}(1+u)Q_{2}(\frac{1}{2}+u)}
 \nonumber \\ 
  &-& \phi(-\frac{3}{2}+ u)\phi(\frac{1}{2}+ u)
  \frac{Q_{1}(-1 + u)Q_{2}(-\frac{1}{2}+ u)Q_{3}(\frac{3}{2}+u)}
       {Q_{1}(u)Q_{2}(\frac{1}{2}+u)Q_{3}(-\frac{1}{2}+u)}
 \nonumber \\ 
 &-& \phi(-\frac{3}{2} + u)\phi(\frac{1}{2}+ u)
       \frac{Q_{1}(-1 + u)Q_{2}(-\frac{1}{2}+u)Q_{3}(-\frac{5}{2}+ u)}
            {Q_{1}(u)Q_{2}(-\frac{3}{2}+ u)Q_{3}(-\frac{1}{2}+u)}
       \nonumber \\ 
  &-& \phi(-\frac{3}{2} + u)\phi(\frac{1}{2} + u)
    \frac{Q_{1}(-1 + u)^2 Q_{2}(-\frac{5}{2} + u)}
    {Q_{1}(-2 + u)Q_{1}(u)Q_{2}(-\frac{3}{2} + u)}
       \nonumber \\
 &+&  \phi(-\frac{5}{2} + u)\phi(\frac{1}{2} + u)
  \frac{Q_{1}(-1 + u)^2}
    {Q_{1}(-2 + u)Q_{1}(u)}
    \nonumber \\ 
 &+& \phi(-\frac{3}{2} + u)\phi(-\frac{1}{2} + u)
\frac{Q_{1}(-1 + u)Q_{3}(\frac{3}{2} + u)}
{Q_{1}(u)Q_{3}(-\frac{1}{2} + u)}
      \nonumber \\
 &+& \phi(-\frac{3}{2} + u)\phi(-\frac{1}{2} + u)
  \frac{Q_{1}(-1 + u)Q_{2}(\frac{1}{2} + u)Q_{3}(-\frac{5}{2} + u)}
       {Q_{1}(u)Q_{2}(-\frac{3}{2} + u)Q_{3}(-\frac{1}{2} + u)}
      \nonumber \\ 
      &+&
      \phi(-\frac{3}{2} + u)\phi(-\frac{1}{2} + u)
      \frac{Q_{1}(-1 + u)^2 Q_{2}(-\frac{5}{2} + u)Q_{2}(\frac{1}{2} + u)}
           {Q_{1}(-2 + u)Q_{1}(u)Q_{2}(-\frac{3}{2} + u)
             Q_{2}(-\frac{1}{2} + u)}
      \nonumber \\
  &-&\phi(-\frac{5}{2} + u)\phi(-\frac{1}{2} + u)
   \frac{Q_{1}(-1 + u)^2 Q_{2}(\frac{1}{2} + u)}
        {Q_{1}(-2 + u)Q_{1}(u)Q_{2}(-\frac{1}{2} + u)}
       \nonumber \\
 &+&  \phi(-\frac{3}{2} + u)\phi(-\frac{1}{2} + u)
  \frac{Q_{1}(-1 + u)
      Q_{2}(-\frac{5}{2} + u)Q_{3}(\frac{1}{2} + u)}{Q_{1}(-2 + u)
      Q_{2}(-\frac{1}{2} + u)Q_{3}(-\frac{3}{2} + u)}
      \nonumber \\
  &-& \phi(-\frac{5}{2} + u)\phi(-\frac{1}{2} + u)
   \frac{Q_{1}(-1 + u)
       Q_{2}(-\frac{3}{2} + u)Q_{3}(\frac{1}{2} + u)}{Q_{1}(-2 + u)
       Q_{2}(-\frac{1}{2} + u)Q_{3}(-\frac{3}{2} + u)}
       \nonumber \\ 
   &+& 
  \phi(-\frac{3}{2} + u)\phi(-\frac{1}{2} + u)
  \frac{Q_{1}(-1 + u)Q_{3}(-\frac{7}{2} + u)}
       {Q_{1}(-2 + u)Q_{3}(-\frac{3}{2} + u)}
      \nonumber \\
  &-& 
  \phi(-\frac{5}{2} + u)\phi(-\frac{1}{2} + u)
     \frac{Q_{1}(-1 + u)
       Q_{2}(-\frac{3}{2} + u)Q_{3}(-\frac{7}{2} + u)}{Q_{1}(-2 + u)
       Q_{2}(-\frac{5}{2} + u)Q_{3}(-\frac{3}{2} + u)}
       \nonumber \\
  &-& \phi(-\frac{5}{2} + u)\phi(-\frac{1}{2} + u)\frac{Q_{1}(-1 + u)
       Q_{2}(-\frac{7}{2} + u)}{Q_{1}(-3 + u)Q_{2}(-\frac{5}{2} + u)}
       \nonumber \\ 
       &+& 
  \phi(-\frac{7}{2} + u)\phi(-\frac{1}{2} + u)
  \frac{Q_{1}(-1 + u)}{Q_{1}(-3 + u)}
  \}
\label{t21-ex}.  
\end{eqnarray}
%
Thanks to Theorem \ref{th-tate} and Theorem \ref{th-yoko}
 (see later),
 these DVFs are free of 
poles under the following BAE: 
\begin{eqnarray}
 && \frac{\phi(u_{k}^{(1)}+\frac{1}{2})}{\phi(u_{k}^{(1)}-\frac{1}{2})}
 =\frac{Q_{2}(u_{k}^{(1)}+\frac{1}{2})}
       {Q_{2}(u_{k}^{(1)}-\frac{1}{2})} \quad {\rm for } \quad 
  1 \le k \le N_{1}, \nonumber \\ 
 && -1=\frac{Q_{1}(u_{k}^{(2)}+\frac{1}{2})
       Q_{2}(u_{k}^{(2)}-1)Q_{3}(u_{k}^{(2)}+1)}
         {Q_{1}(u_{k}^{(2)}-\frac{1}{2})
         Q_{2}(u_{k}^{(2)}+1)Q_{3}(u_{k}^{(2)}-1)} 
         \quad {\rm for } \quad 
  1 \le k \le N_{2}, \nonumber \\ 
 && -1=\frac{Q_{2}(u_{k}^{(3)}+1)Q_{3}(u_{k}^{(3)}-2)}
         {Q_{2}(u_{k}^{(3)}-1)Q_{3}(u_{k}^{(3)}+2)},
          \quad {\rm for } \quad  
    1 \le k \le N_{3}.
\label{BAE2}
\end{eqnarray}
We note that DVFs have so called 
{\em Bethe-strap} structures 
\cite{KS1,S2}, which bear resemblance to 
weight space diagrams. 
 For example, Bethe-strap structures of (\ref{t1-ex}), 
(\ref{t2-ex}) and (\ref{t21-ex}) are given in 
Figure \ref{best1}, Figure \ref{best2} and Figure \ref{best3} 
respectively.  
We also note that the Bethe-stap resembles crystal graph 
\cite{KN1,KN2} as was pointed out in \cite{KS1} for simple 
 Lie algebras cases.

{\em Remark}: 
Recently we have found curious terms in 
many Bethe-straps. They may be called 
{\em pseudo-top terms}, 
which have the following properties 
(cf. Figure \ref{best1}, Figure \ref{best2}, Figure \ref{best3}): 
\begin{enumerate}
  \item 
  
  They carry Lie (super) algebras weights, which are 
  lower than the one for the top term. 
  
 \item 

	They send out arrows but does not suck in arrows.
\end{enumerate} 
We also found {\em pseudo-bottom terms}, which 
have the following properties: 
\begin{enumerate}
  \item 
  
  They carry Lie (super) algebras weights, which are 
  higher than the one for the bottom term. 
  
	\item 
	They  suck in arrows but does not send out arrows.
\end{enumerate} 
 We found a pseudo-top term in the Bethe-strap of
  $T_{2}^{(3)}(u)$ (As for $T_{m}^{(a)}(u)$, see section 5.) 
  for $C(4)$. 
  Pseudo-top term and pseudo-bottom term 
  are not peculiar for Bethe-straps 
  related to Lie superalgebras. 
 Actually, we found such terms even for $sl_{2}$ case: 
 the DVF $T_{2}(u)T_{1}(u+k)$ has a pseudo-top 
 term if $k=1$; a pseudo-bottom 
 term if $k=-1$
 (see \cite{KS1} for the definition of $T_{m}(u)$). 
 In spite of existence of these curious terms, 
 auxiliary spaces of DVFs in this paper are 
 supposed to correspond to 
 some {\em irreducible} representations as long as 
 the Bethe straps are {\em connected}  in the whole 
 (cf \cite{KS1,KOS,K2}). 
  In fact we have confirmed, for several cases, 
  that the Bethe straps of the DVFs ${\cal T}^{a}(u)$ and 
  $T_{m}^{(a)}(u)$  are connected in the whole.  
We hope to discuss about these curious 
terms in detail elsewhere. 
\begin{figure}
    \setlength{\unitlength}{1.5pt}
    \begin{center}
    \begin{picture}(250,310) 
      \put(120,0){\line(1,0){10}}
      \put(130,0){\line(0,1){20}}
      \put(130,20){\line(-1,0){10}}
      \put(120,20){\line(0,-1){20}}
      \put(120,10){\line(1,0){10}}
      \put(123,12){$\overline{2}$}
      \put(123,2){$\overline{1}$}
      \put(114,9){$-$}
      \put(80,40){\line(1,0){10}}
      \put(90,40){\line(0,1){20}}
      \put(90,60){\line(-1,0){10}}
      \put(80,60){\line(0,-1){20}}
      \put(80,50){\line(1,0){10}}
      \put(83,52){$\overline{2}$}
      \put(83,42){$\overline{2}$}
      \put(160,40){\line(1,0){10}}
      \put(170,40){\line(0,1){20}}
      \put(170,60){\line(-1,0){10}}
      \put(160,60){\line(0,-1){20}}
      \put(160,50){\line(1,0){10}}
      \put(163,52){$\overline{3}$}
      \put(163,42){$\overline{1}$}
      \put(154,49){$-$}
      \put(80,80){\line(1,0){10}}
      \put(90,80){\line(0,1){20}}
      \put(90,100){\line(-1,0){10}}
      \put(80,100){\line(0,-1){20}}
      \put(80,90){\line(1,0){10}}
      \put(83,92){$\overline{3}$}
      \put(83,82){$\overline{2}$}
      \put(160,80){\line(1,0){10}}
      \put(170,80){\line(0,1){20}}
      \put(170,100){\line(-1,0){10}}
      \put(160,100){\line(0,-1){20}}
      \put(160,90){\line(1,0){10}}
      \put(163,92){$3$}
      \put(163,82){$\overline{1}$}
      \put(154,89){$-$}
      \put(40,120){\line(1,0){10}}
      \put(50,120){\line(0,1){20}}
      \put(50,140){\line(-1,0){10}}
      \put(40,140){\line(0,-1){20}}
      \put(40,130){\line(1,0){10}}
      \put(43,132){$\overline{3}$}
      \put(43,122){$\overline{3}$}
      \put(120,120){\line(1,0){10}}
      \put(130,120){\line(0,1){20}}
      \put(130,140){\line(-1,0){10}}
      \put(120,140){\line(0,-1){20}}
      \put(120,130){\line(1,0){10}}
      \put(123,132){$3$}
      \put(123,122){$\overline{2}$}
      \put(200,120){\line(1,0){10}}
      \put(210,120){\line(0,1){20}}
      \put(210,140){\line(-1,0){10}}
      \put(200,140){\line(0,-1){20}}
      \put(200,130){\line(1,0){10}}
      \put(203,132){$2$}
      \put(203,122){$\overline{1}$}
      \put(194,129){$-$}
      \put(0,160){\line(1,0){10}}
      \put(10,160){\line(0,1){20}}
      \put(10,180){\line(-1,0){10}}
      \put(0,180){\line(0,-1){20}}
      \put(0,170){\line(1,0){10}}
      \put(3,172){$\overline{3}$}
      \put(3,162){$3$}
      \put(80,160){\line(1,0){10}}
      \put(90,160){\line(0,1){20}}
      \put(90,180){\line(-1,0){10}}
      \put(80,180){\line(0,-1){20}}
      \put(80,170){\line(1,0){10}}
      \put(83,172){$3$}
      \put(83,162){$\overline{3}$}
      \put(160,160){\line(1,0){10}}
      \put(170,160){\line(0,1){20}}
      \put(170,180){\line(-1,0){10}}
      \put(160,180){\line(0,-1){20}}
      \put(160,170){\line(1,0){10}}
      \put(163,172){$2$}
      \put(163,162){$\overline{2}$}
      \put(240,160){\line(1,0){10}}
      \put(250,160){\line(0,1){20}}
      \put(250,180){\line(-1,0){10}}
      \put(240,180){\line(0,-1){20}}
      \put(240,170){\line(1,0){10}}
      \put(243,172){$1$}
      \put(243,162){$\overline{1}$}
      \put(40,200){\line(1,0){10}}
      \put(50,200){\line(0,1){20}}
      \put(50,220){\line(-1,0){10}}
      \put(40,220){\line(0,-1){20}}
      \put(40,210){\line(1,0){10}}
      \put(43,212){$3$}
      \put(43,202){$3$}
      \put(120,200){\line(1,0){10}}
      \put(130,200){\line(0,1){20}}
      \put(130,220){\line(-1,0){10}}
      \put(120,220){\line(0,-1){20}}
      \put(120,210){\line(1,0){10}}
      \put(123,212){$2$}
      \put(123,202){$\overline{3}$}
      \put(200,200){\line(1,0){10}}
      \put(210,200){\line(0,1){20}}
      \put(210,220){\line(-1,0){10}}
      \put(200,220){\line(0,-1){20}}
      \put(200,210){\line(1,0){10}}
      \put(203,212){$1$}
      \put(203,202){$\overline{2}$}
      \put(194,209){$-$}
      \put(80,240){\line(1,0){10}}
      \put(90,240){\line(0,1){20}}
      \put(90,260){\line(-1,0){10}}
      \put(80,260){\line(0,-1){20}}
      \put(80,250){\line(1,0){10}}
      \put(83,252){$2$}
      \put(83,242){$3$}
      \put(160,240){\line(1,0){10}}
      \put(170,240){\line(0,1){20}}
      \put(170,260){\line(-1,0){10}}
      \put(160,260){\line(0,-1){20}}
      \put(160,250){\line(1,0){10}}
      \put(163,252){$1$}
      \put(163,242){$\overline{3}$}
      \put(154,249){$-$}
      \put(80,280){\line(1,0){10}}
      \put(90,280){\line(0,1){20}}
      \put(90,300){\line(-1,0){10}}
      \put(80,300){\line(0,-1){20}}
      \put(80,290){\line(1,0){10}}
      \put(83,292){$2$}
      \put(83,282){$2$}
      \put(160,280){\line(1,0){10}}
      \put(170,280){\line(0,1){20}}
      \put(170,300){\line(-1,0){10}}
      \put(160,300){\line(0,-1){20}}
      \put(160,290){\line(1,0){10}}
      \put(163,292){$1$}
      \put(163,282){$3$}
      \put(154,289){$-$}
      \put(120,320){\line(1,0){10}}
      \put(130,320){\line(0,1){20}}
      \put(130,340){\line(-1,0){10}}
      \put(120,340){\line(0,-1){20}}
      \put(120,330){\line(1,0){10}}
      \put(123,332){$1$}
      \put(123,322){$2$}
      \put(114,329){$-$}
      %
      \put(92,38){\vector(3,-2){26}}
      \put(102,33){$\scriptsize{(1,3)}$}
      \put(52,118){\vector(3,-2){26}}
      \put(62,113){$\scriptsize{(2,\frac{5}{2})}$}
      \put(132,118){\vector(3,-2){26}}
      \put(142,113){$\scriptsize{(1,3)}$}
      \put(12,158){\vector(3,-2){26}}
      \put(22,153){$\scriptsize{(3,\frac{3}{2})}$}
      \put(92,158){\vector(3,-2){26}}
      \put(102,153){$\scriptsize{(2,\frac{5}{2})}$}
      \put(172,158){\vector(3,-2){26}}
      \put(182,153){$\scriptsize{(1,3)}$}
      \put(52,198){\vector(3,-2){26}}
      \put(62,193){$\scriptsize{(3,\frac{3}{2})}$}
      \put(132,198){\vector(3,-2){26}}
      \put(142,193){$\scriptsize{(2,\frac{5}{2})}$}
      \put(212,198){\vector(3,-2){26}}
      \put(222,193){$\scriptsize{(1,3)}$}
      \put(172,238){\vector(3,-2){26}}
      \put(182,233){$\scriptsize{(2,\frac{5}{2})}$}
      \put(92,238){\vector(3,-2){26}}
      \put(102,233){$\scriptsize{(3,\frac{3}{2})}$}
      \put(132,318){\vector(3,-2){26}}
      \put(142,313){$\scriptsize{(2,\frac{1}{2})}$}
      \put(158,38){\vector(-3,-2){26}}
      \put(132,34){$\scriptsize{(2,\frac{3}{2})}$}
      \put(198,118){\vector(-3,-2){26}}
      \put(169,116){$\scriptsize{(2,-\frac{1}{2})}$}
      \put(118,118){\vector(-3,-2){26}}
      \put(92,115){$\scriptsize{(3,\frac{1}{2})}$}
      \put(238,158){\vector(-3,-2){26}}
      \put(208,155){$\scriptsize{(1,-1)}$}
      \put(158,158){\vector(-3,-2){26}}
      \put(129,156){$\scriptsize{(2,-\frac{1}{2})}$}
      \put(78,158){\vector(-3,-2){26}}
      \put(52,155){$\scriptsize{(3,\frac{1}{2})}$}
      \put(38,198){\vector(-3,-2){26}}
      \put(13,195){$\scriptsize{(3,\frac{1}{2})}$}
      \put(118,198){\vector(-3,-2){26}}
      \put(88,195){$\scriptsize{(2,-\frac{1}{2})}$}
      \put(198,198){\vector(-3,-2){26}}
      \put(168,195){$\scriptsize{(1,-1)}$}
      \put(78,238){\vector(-3,-2){26}}
      \put(48,236){$\scriptsize{(2,-\frac{1}{2})}$}
      \put(158,238){\vector(-3,-2){26}}
      \put(126,235){$\scriptsize{(1,-1)}$}
      \put(118,318){\vector(-3,-2){26}}
      \put(84,312){$\scriptsize{(1,-1)}$}
      %
      \put(85,77){\vector(0,-1){14}}
      \put(87,69){$\scriptsize{(2,\frac{3}{2})}$}
      \put(165,77){\vector(0,-1){14}}
      \put(167,69){$\scriptsize{(3,\frac{1}{2})}$}
      \put(85,277){\vector(0,-1){14}}
      \put(87,269){$\scriptsize{(2,\frac{1}{2})}$}
      \put(165,277){\vector(0,-1){14}}
      \put(167,269){$\scriptsize{(3,\frac{3}{2})}$}
      \put(94,80){\vector(3,-1){62}}
      \put(124,72){$\scriptsize{(1,3)}$}
      \put(156,280){\vector(-3,-1){62}}
      \put(116,277){$\scriptsize{(1,-1)}$}
  \end{picture}
  \end{center}
  \caption{The Bethe-strap structure of the 
  DVF ${\cal T}^{2}(u)$ (\ref{t2-ex}) for 
   the Lie superalgebra $C(3)$:  
 The topmost (resp. bottommost) tableau corresponds to the 
 \symbol{96}highest weight vector \symbol{39} 
 (resp. \symbol{96}lowest weight vector \symbol{39}),
 which is called the {\em top term} 
 (resp. {\em bottom term }).}
  \label{best2}
\end{figure}
\begin{figure}
    \setlength{\unitlength}{1.5pt}
    \begin{center}
    \begin{picture}(100,310) 
      \put(40,0){\line(1,0){20}}
      \put(60,0){\line(0,1){10}}
      \put(60,10){\line(-1,0){20}}
      \put(40,10){\line(0,-1){10}}
      \put(50,0){\line(0,1){10}}
      \put(43,2){$\overline{1}$}
      \put(53,2){$\overline{1}$}
      \put(40,30){\line(1,0){20}}
      \put(60,30){\line(0,1){10}}
      \put(60,40){\line(-1,0){20}}
      \put(40,40){\line(0,-1){10}}
      \put(50,30){\line(0,1){10}}
      \put(43,32){$\overline{2}$}
      \put(53,32){$\overline{1}$}
      \put(34,33){$-$}
      \put(40,60){\line(1,0){20}}
      \put(60,60){\line(0,1){10}}
      \put(60,70){\line(-1,0){20}}
      \put(40,70){\line(0,-1){10}}
      \put(50,60){\line(0,1){10}}
      \put(43,62){$\overline{3}$}
      \put(53,62){$\overline{1}$}
      \put(34,63){$-$}
      \put(0,90){\line(1,0){20}}
      \put(20,90){\line(0,1){10}}
      \put(20,100){\line(-1,0){20}}
      \put(0,100){\line(0,-1){10}}
      \put(10,90){\line(0,1){10}}
      \put(3,92){$\overline{3}$}
      \put(13,92){$\overline{2}$}
      \put(80,90){\line(1,0){20}}
      \put(100,90){\line(0,1){10}}
      \put(100,100){\line(-1,0){20}}
      \put(80,100){\line(0,-1){10}}
      \put(90,90){\line(0,1){10}}
      \put(83,92){$3$}
      \put(93,92){$\overline{1}$}
      \put(74,93){$-$}
      \put(0,120){\line(1,0){20}}
      \put(20,120){\line(0,1){10}}
      \put(20,130){\line(-1,0){20}}
      \put(0,130){\line(0,-1){10}}
      \put(10,120){\line(0,1){10}}
      \put(3,122){$3$}
      \put(13,122){$\overline{2}$}
      \put(80,120){\line(1,0){20}}
      \put(100,120){\line(0,1){10}}
      \put(100,130){\line(-1,0){20}}
      \put(80,130){\line(0,-1){10}}
      \put(90,120){\line(0,1){10}}
      \put(83,122){$2$}
      \put(93,122){$\overline{1}$}
      \put(74,123){$-$}
      \put(0,150){\line(1,0){20}}
      \put(20,150){\line(0,1){10}}
      \put(20,160){\line(-1,0){20}}
      \put(0,160){\line(0,-1){10}}
      \put(10,150){\line(0,1){10}}
      \put(3,152){$2$}
      \put(13,152){$\overline{2}$}
      \put(80,150){\line(1,0){20}}
      \put(100,150){\line(0,1){10}}
      \put(100,160){\line(-1,0){20}}
      \put(80,160){\line(0,-1){10}}
      \put(90,150){\line(0,1){10}}
      \put(83,152){$1$}
      \put(93,152){$\overline{1}$}
      \put(0,180){\line(1,0){20}}
      \put(20,180){\line(0,1){10}}
      \put(20,190){\line(-1,0){20}}
      \put(0,190){\line(0,-1){10}}
      \put(10,180){\line(0,1){10}}
      \put(3,182){$2$}
      \put(13,182){$\overline{3}$}
      \put(80,180){\line(1,0){20}}
      \put(100,180){\line(0,1){10}}
      \put(100,190){\line(-1,0){20}}
      \put(80,190){\line(0,-1){10}}
      \put(90,180){\line(0,1){10}}
      \put(83,182){$1$}
      \put(93,182){$\overline{2}$}
      \put(74,183){$-$}
      \put(0,210){\line(1,0){20}}
      \put(20,210){\line(0,1){10}}
      \put(20,220){\line(-1,0){20}}
      \put(0,220){\line(0,-1){10}}
      \put(10,210){\line(0,1){10}}
      \put(3,212){$2$}
      \put(13,212){$3$}
      \put(80,210){\line(1,0){20}}
      \put(100,210){\line(0,1){10}}
      \put(100,220){\line(-1,0){20}}
      \put(80,220){\line(0,-1){10}}
      \put(90,210){\line(0,1){10}}
      \put(83,212){$1$}
      \put(93,212){$\overline{3}$}
      \put(74,213){$-$}
      \put(40,240){\line(1,0){20}}
      \put(60,240){\line(0,1){10}}
      \put(60,250){\line(-1,0){20}}
      \put(40,250){\line(0,-1){10}}
      \put(50,240){\line(0,1){10}}
      \put(43,242){$1$}
      \put(53,242){$3$}
      \put(34,243){$-$}
      \put(40,270){\line(1,0){20}}
      \put(60,270){\line(0,1){10}}
      \put(60,280){\line(-1,0){20}}
      \put(40,280){\line(0,-1){10}}
      \put(50,270){\line(0,1){10}}
      \put(43,272){$1$}
      \put(53,272){$2$}
      \put(34,273){$-$}
      \put(40,300){\line(1,0){20}}
      \put(60,300){\line(0,1){10}}
      \put(60,310){\line(-1,0){20}}
      \put(40,310){\line(0,-1){10}}
      \put(50,300){\line(0,1){10}}
      \put(43,302){$1$}
      \put(53,302){$1$}
      \put(50,28){\vector(0,-1){16}}
      \put(52,19){$\scriptsize{(1,3)}$}
      \put(50,58){\vector(0,-1){16}}
      \put(52,49){$\scriptsize{(2,\frac{5}{2})}$}
      \put(10,118){\vector(0,-1){16}}
      \put(12,109){$\scriptsize{(3,\frac{3}{2})}$}
      \put(90,118){\vector(0,-1){16}}
      \put(92,109){$\scriptsize{(2,\frac{1}{2})}$}
      \put(10,148){\vector(0,-1){16}}
      \put(12,139){$\scriptsize{(2,\frac{1}{2})}$}
      \put(90,148){\vector(0,-1){16}}
      \put(92,139){$\scriptsize{(1,0)}$}
      \put(10,178){\vector(0,-1){16}}
      \put(12,169){$\scriptsize{(2,\frac{3}{2})}$}
      \put(90,178){\vector(0,-1){16}}
      \put(92,169){$\scriptsize{(1,2)}$}
      \put(10,208){\vector(0,-1){16}}
      \put(12,199){$\scriptsize{(3,\frac{1}{2})}$}
      \put(90,208){\vector(0,-1){16}}
      \put(92,199){$\scriptsize{(2,\frac{3}{2})}$}
      \put(50,268){\vector(0,-1){16}}
      \put(52,259){$\scriptsize{(2,-\frac{1}{2})}$}
      \put(50,298){\vector(0,-1){16}}
      \put(52,289){$\scriptsize{(1,-1)}$}
      \put(22,88){\vector(1,-1){16}}
      \put(31,81){$\scriptsize{(1,2)}$}
      \put(62,238){\vector(1,-1){16}}
      \put(71,231){$\scriptsize{(3,\frac{1}{2})}$}
      \put(78,88){\vector(-1,-1){16}}
      \put(53,83){$\scriptsize{(3,\frac{3}{2})}$}
      \put(38,238){\vector(-1,-1){16}}
      \put(14,233){$\scriptsize{(1,0)}$}
      \put(22,119){\vector(3,-1){55}}
      \put(46,114){$\scriptsize{(1,2)}$}
      \put(22,149){\vector(3,-1){55}}
      \put(46,144){$\scriptsize{(1,2)}$}
      \put(78,179){\vector(-3,-1){55}}
      \put(46,176){$\scriptsize{(1,0)}$}
      \put(78,209){\vector(-3,-1){55}}
      \put(46,206){$\scriptsize{(1,0)}$}
  \end{picture}
  \end{center}
  \caption{The Bethe-strap structure of the 
  DVF $T_{2}^{(1)}(u)$ (\ref{t21-ex}) for 
   the Lie superalgebra $C(3)$:  
 The topmost (resp. bottommost) tableau corresponds to the 
 \symbol{96}highest weight vector \symbol{39}
 (resp. \symbol{96}lowest weight vector \symbol{39}), 
 which is called the {\em top term} 
 (resp. {\em bottom term}).}
  \label{best3}
\end{figure}
 
In general, we have the following Theorems, which
 are essential in the analytic Bethe ansatz.
\begin{theorem} \label{th-tate}
For  $a\in {\bf Z}_{\ge 0}$, DVF ${\cal T}^{a}(u)$  (\ref{tate}) 
is free of poles under the condition that
the BAE {\rm (\ref{BAE})} is valid
\footnote{We assume that   
$u_{i}^{(b)}-u_{j}^{(b)} \ne (\alpha_{b}|\alpha_{b})$
 for any $i,j \in \{1,2,\dots , N_{b}\} $ ($i \ne j$) 
 and $b\in \{1,2,\dots, s \}$ 
 in BAE (\ref{BAE}). 
 If this assumption does not hold, we will 
 need separate consideration. 
 This assumption may require detailed analysis of the 
 BAE (\ref{BAE}), which is beyond the scope of this paper. 
 We also note that similar assumptions are implicitly 
 assumed in many literature concerning 
 analytic Bethe ansatz. }
\end{theorem}
\begin{theorem}\label{th-yoko} 
For $m\in \{0,1,2,\dots, s-1\}$,
 DVF $T_{m}^{(1)}(u)$  (\ref{yoko}) 
is free of poles under the condition that
the BAE {\rm (\ref{BAE})} is valid.   
\end{theorem} 
We can prove Theorem \ref{th-yoko} by using the following lemmas. 
\begin{lemma}
(1) For  $b \in \{2,3,\dots,s-1 \}$,  
\begin{eqnarray}
&& Res_{v=u_{k}^{(b)}+\frac{b}{2}}( 
\framebox{$b$}_{v}\framebox{$\overline{b+1}$}_{v+s-b}+
\framebox{$b$}_{v}\framebox{$\overline{b}$}_{v+s-b}
)=0, \\ 
&& Res_{v=u_{k}^{(b)}+\frac{b-2}{2}}(
\framebox{$b$}_{v}\framebox{$\overline{b}$}_{v+s-b}+
\framebox{$b+1$}_{v}\framebox{$\overline{b}$}_{v+s-b})=0
\end{eqnarray}
 under the BAE {\rm (\ref{BAE})}.   \\ 
(2)  
\begin{equation}
\framebox{$b$}_{v}\framebox{$\overline{b+1}$}_{v+s-b}+
\framebox{$b$}_{v}\framebox{$\overline{b}$}_{v+s-b}+
\framebox{$b+1$}_{v}\framebox{$\overline{b}$}_{v+s-b}
\end{equation}
is free of color $b$ 
($b \in \{2,3,\dots,s-1 \}$)
 pole under the BAE {\rm (\ref{BAE})}.
\end{lemma}
The following lemma is $C(s)$ version of lemma 3.3.4. 
in \cite{KS1} for 
$C_{s}$.
\begin{lemma} \label{le-yoko}
For  $b\in \{2,3,\dots,s-1\}$, let  
\begin{eqnarray}
  \begin{array}{|c|c|c|c|c|}\hline
    \xi & b & \eta & \overline{b+1} & \zeta  \\
    \hline 
  \end{array},\quad 
  \begin{array}{|c|c|c|c|c|}\hline
    \xi & b+1 & \eta & \overline{b} & \zeta  \\
    \hline 
  \end{array},
\end{eqnarray}
be the terms that appear in $T_{m}^{(1)}(u)$ (\ref{yoko}).
Here \framebox{$\xi $}, 
\framebox{$\eta $} and \framebox{$\zeta $}  
 do not contain \framebox{$b$},\framebox{$b+1$},
 \framebox{$\overline{b+1}$}, \framebox{$\overline{b}$}.
 In this case, the length of \framebox{$\eta $}  is less than $s-b$.
\end{lemma}
We can prove  Theorem \ref{th-yoko} by a similar idea used in 
the proof of  
Theorem 3.3.1. in \cite{KS1}. So we prove only Theorem \ref{th-tate} 
from now on.  

{\em Proof of Theorem \ref{th-tate}}. 
For simplicity, we assume that the vacuum parts are formally trivial, 
that is, the left hand side of the 
 BAE (\ref{BAE}) is constantly $-1$. 
We prove that ${\cal T}^{a}(u)$ (\ref{tate}) 
is free of color $b$ pole, namely, 
$Res_{u=u_{k}^{(b)}+\cdots}{\cal T}^{a}(u)=0$ for any 
 $ b \in \{1,2,\dots, s\} $
 under the condition that the BAE (\ref{BAE}) is valid. 
 The function $\framebox{$c$}_{u}$  
 (\ref{z+}) with $c \in J $ has 
 color $b$ poles only at $c=b$, $b+1$, 
$\overline{b+1}$ or $\overline{b}$ for 
$b\in \{1,2,\dots,s-1 \} $; 
at $c=s$ or $c=\overline{s}$ for $b=s$, so we shall trace only 
\framebox{$b$}, \framebox{$b+1$}, 
\framebox{$\overline{b+1}$} or \framebox{$\overline{b}$}
 for $b\in\{1,2,\dots, s-1 \}$; 
\framebox{$s$} or \framebox{$\overline{s}$} for $b=s$.
Denote $S_{k}$ the partial sum of ${\cal T}^{a}(u)$, which contains 
$k$ boxes among \framebox{$b$}, \framebox{$b+1$}, 
\framebox{$\overline{b+1}$} or \framebox{$\overline{b}$}
 for $1\le b \le s-1$; 
\framebox{$s$} or \framebox{$\overline{s}$} for $b=s$.
 Apparently, $S_{0}$ does not have 
color $b$ pole. 

 Now we examine $S_{1}$ which is a summation  
 of the tableaux (with sign) of the form 
\begin{equation} 
\begin{array}{|c|}\hline
    \xi \\ \hline 
    \eta   \\ \hline 
   \zeta \\ \hline
\end{array}
\label{tableaux}
\end{equation}
where \framebox{$\xi$} and \framebox{$\zeta$} are columns with 
total length $a-1$ and they do not involve $Q_{b}$. 
\framebox{$\eta$} is \framebox{$b$}, \framebox{$b+1$}, 
\framebox{$\overline{b+1}$} or \framebox{$\overline{b}$}
 for $1\le b \le s-1$; 
\framebox{$s$} or \framebox{$\overline{s}$} for $b=s$. 
 Thanks to the relations (\ref{res1})-(\ref{res5}), 
 $S_{1}$ is free of color $b$ pole under the BAE (\ref{BAE}). 
 Hereafter we consider $S_{k}$  ($k\ge 2$). \\ 
$\bullet$ 
The case $ b=1 $ : $S_{k} (k\ge 2)$ is a 
summation of the tableaux (with sign) of the form 
\begin{eqnarray}
&& \begin{array}{|c|}\hline 
    D_{11}    \\ \hline 
    \eta \\ \hline 
    D_{21}  \\ \hline
\end{array}
-\begin{array}{|c|}\hline 
    D_{11}    \\ \hline 
    \eta  \\ \hline 
    D_{22}  \\ \hline 
\end{array}
-\begin{array}{|c|}\hline
    D_{12}   \\ \hline 
    \eta  \\ \hline 
    D_{21}  \\ \hline
\end{array}
+\begin{array}{|c|}\hline 
    D_{12}  \\ \hline 
    \eta  \\ \hline 
    D_{22}  \\ \hline 
\end{array} \nonumber \\ 
&&=(
\begin{array}{|c|}\hline 
    D_{11}  \\ \hline 
\end{array}
-\begin{array}{|c|}\hline 
    D_{12}  \\ \hline 
\end{array}
)(
\begin{array}{|c|}\hline 
    D_{21}  \\ \hline 
\end{array}
-\begin{array}{|c|}\hline 
    D_{22}  \\ \hline 
\end{array}
)
\begin{array}{|c|}\hline 
    \eta \\ \hline 
\end{array}
\label{tableauxk1}
\end{eqnarray} 
where \framebox{$\eta$} is a column with 
length $a-k$, which does not contain \framebox{$1$}, 
\framebox{$2$}, \framebox{$\overline{2}$} and 
\framebox{$\overline{1}$}; 
\framebox{$D_{11}$} is a column 
\footnote{We assume that 
$\framebox{$D_{11}$}=\framebox{$1$}_{v}$ if $k_{1}=1$.}
of the form: 
\begin{equation}
\begin{array}{|c|l}\cline{1-1}
    1 & _v  \\ \cline{1-1} 
    2 & _{v-1} \\ \cline{1-1} 
   \vdots & \\  \cline{1-1} 
    2 & _{v-k_{1}+1} \\ \cline{1-1} 
\end{array}
= \frac{Q_{1}(v+\frac{1}{2}-k_{1})Q_{2}(v)}
       {Q_{1}(v+\frac{1}{2})Q_{2}(v-k_{1}+1)}; 
\label{tableauxk1-1}
\end{equation}
\framebox{$D_{12}$} is a column 
of the form:  
\begin{equation}
\begin{array}{|c|l}\cline{1-1}
    2 & _v \\ \cline{1-1} 
    2 & _{v-1} \\ \cline{1-1} 
   \vdots & \\ \cline{1-1} 
    2 & _{v-k_{1}+1}\\ \cline{1-1} 
\end{array}
=\frac{Q_{1}(v+\frac{1}{2}-k_{1})Q_{2}(v+1)}
       {Q_{1}(v+\frac{1}{2})Q_{2}(v-k_{1}+1)}
\label{tableauxk1-2}, 
\end{equation}
where $v=u+h_{1}$: $h_{1}$ is some shift parameter; 
\framebox{$D_{21}$} is a column
\footnote{We assume that 
$\framebox{$D_{21}$}=\framebox{$\overline{1}$}_{w}$
 if $k_{2}=1$.}
 of the form: 
\begin{equation}
\begin{array}{|c|l}\cline{1-1}
    \overline{2} & _w  \\ \cline{1-1} 
    \vdots & \\  \cline{1-1} 
    \overline{2} & _{w-k_{2}+2} \\ \cline{1-1}
    \overline{1} & _{w-k_{2}+1} \\ \cline{1-1} 
\end{array}
= \frac{Q_{1}(w-\frac{2s-3}{2})Q_{2}(w-\frac{2s-4}{2}-k_{2})}
       {Q_{1}(w-\frac{2s-3}{2}-k_{2})Q_{2}(w-\frac{2s-2}{2})} ;
\label{tableauxk1-3}
\end{equation}
\framebox{$D_{22}$} is a column of the form:
\begin{equation}
\begin{array}{|c|l}\cline{1-1}
    \overline{2} & _w \\ \cline{1-1} 
    \vdots & \\ \cline{1-1} 
    \overline{2} & _{w-k_{2}+2} \\ \cline{1-1}
    \overline{2} & _{w-k_{2}+1}\\ \cline{1-1} 
\end{array}
=\frac{Q_{1}(w-\frac{2s-3}{2})Q_{2}(w-\frac{2s-2}{2}-k_{2})}
      {Q_{1}(w-\frac{2s-3}{2}-k_{2})Q_{2}(w-\frac{2s-2}{2})}
       \label{tableauxk1-4}, 
\end{equation}
where $w=u+h_{2}$: $h_{2}$ is some shift parameter; 
$k=k_{1}+k_{2}$
 \footnote{Here we discussed the case for 
 $k_{1}\ge 1 $ and $k_{2}\ge 1 $; 
 the case for $k_{1}=0 $ or $k_{2}=0 $ can be treated similarly.}. 
Obviously, the color $b=1$ residues at $v=-\frac{1}{2}+u_{j}^{(1)}$ 
 in (\ref{tableauxk1-1}) and 
 (\ref{tableauxk1-2}) cancel each other
 under the BAE (\ref{BAE}). 
And the color $b=1$ residues
 at $w=\frac{2s-3}{2}+k_{2}+u_{j}^{(1)}$ 
 in (\ref{tableauxk1-3}) and
 (\ref{tableauxk1-4}) cancel each other
 under the BAE (\ref{BAE}).  
 Thus $S_{k}$ does not 
 have color $1$ pole under the BAE (\ref{BAE}). \\ 
$\bullet$ 
The case $2 \le b \le s-1$: $S_{k} (k\ge 2)$ is a  
summation of the tableaux (with sign) of the form 
\begin{eqnarray}
&& 
\sum_{n_{1}=0}^{k_{1}}
\sum_{n_{2}=0}^{k_{2}} 
\begin{array}{|c|}\hline 
    \xi \\ \hline
    E_{1n_{1}}    \\ \hline 
    \eta \\ \hline 
    E_{2n_{2}}  \\ \hline 
    \zeta \\ \hline 
\end{array}
 \nonumber \\ 
&& =
\left(
\sum_{n_{1}=0}^{k_{1}} 
\begin{array}{|c|}\hline 
    E_{1n_{1}}    \\ \hline 
\end{array}
\right)
\left(
\sum_{n_{2}=0}^{k_{2}}
\begin{array}{|c|}\hline 
    E_{2n_{2}}    \\ \hline 
\end{array}
\right)
\times 
\begin{array}{|c|}\hline 
    \xi \\ \hline 
\end{array}
\times 
\begin{array}{|c|}\hline 
    \eta \\ \hline 
\end{array}
\times 
\begin{array}{|c|}\hline 
    \zeta \\ \hline 
\end{array}
\label{tableauxk2}, 
\end{eqnarray} 
where \framebox{$\xi$}, \framebox{$\eta$} and 
\framebox{$\zeta$} are columns with total 
length $a-k$, which do not contain \framebox{$b$}, 
\framebox{$b+1$}, \framebox{$\overline{b+1}$} and 
\framebox{$\overline{b}$}; 
\framebox{$E_{1n_{1}}$} is a column 
\footnote{We assume that 
$\framebox{$E_{10}$}=
\begin{array}{|c|l}\cline{1-1} 
    b+1 & _{v} \\ \cline{1-1} 
   \vdots & \\ \cline{1-1} 
    b+1 & _{v-k_{1}+1}\\ \cline{1-1} 
 \end{array}
$
 and  
 $\framebox{$E_{1 k_{1}}$}=
\begin{array}{|c|l}\cline{1-1} 
    b & _{v} \\ \cline{1-1} 
   \vdots & \\ \cline{1-1} 
    b & _{v-k_{1}+1}\\ \cline{1-1} 
 \end{array}
$. \\ \ }
of the form: 
\begin{eqnarray}
 \begin{array}{|c|l}\cline{1-1} 
    b   & _v \\ \cline{1-1} 
    \vdots & \\ \cline{1-1} 
    b & _{v-n_{1}+1}\\ \cline{1-1} 
    b+1 & _{v-n_{1}} \\ \cline{1-1} 
   \vdots & \\ \cline{1-1} 
    b+1 & _{v-k_{1}+1}\\ \cline{1-1} 
 \end{array} 
&=&\frac{Q_{b-1}(v-\frac{b-3}{2}-n_{1})Q_{b}(v-\frac{b-4}{2})}
      {Q_{b-1}(v-\frac{b-3}{2})Q_{b}(v-\frac{b-4}{2}-n_{1})}  
\label{tableauxk3} \\ 
&& \times  
 \frac{Q_{b}(v-\frac{b-2}{2}-k_{1})Q_{b+1}(v-\frac{b-3}{2}-n_{1})}
      {Q_{b}(v-\frac{b-2}{2}-n_{1})Q_{b+1}(v-\frac{b-3}{2}-k_{1})}
\nonumber
\end{eqnarray}
for $2\le b \le s-2$;
\begin{eqnarray}
 \begin{array}{|c|l}\cline{1-1} 
    s-1   & _v \\ \cline{1-1} 
    \vdots & \\ \cline{1-1} 
    s-1 & _{v-n_{1}+1}\\ \cline{1-1} 
    s & _{v-n_{1}} \\ \cline{1-1} 
   \vdots & \\ \cline{1-1} 
    s & _{v-k_{1}+1}\\ \cline{1-1} 
 \end{array} 
&=&\frac{Q_{s-2}(v-\frac{s-4}{2}-n_{1})Q_{s-1}(v-\frac{s-5}{2})}
      {Q_{s-2}(v-\frac{s-4}{2})Q_{s-1}(v-\frac{s-5}{2}-n_{1})} 
\label{tableauxk4} \\ 
& \times &  
 \frac{Q_{s-1}(v-\frac{s-3}{2}-k_{1})
          Q_{s}(v-\frac{s-5}{2}-n_{1})Q_{s}(v-\frac{s-3}{2}-n_{1})}
      {Q_{s-1}(v-\frac{s-3}{2}-n_{1})Q_{s}(v-\frac{s-3}{2}-k_{1})
        Q_{s}(v-\frac{s-5}{2}-k_{1})}
\nonumber
\end{eqnarray}
for
\footnote{
We need not take care of the sequence of 
$\begin{array}{|c|}\hline  
    \overline{s}  \\ \hline 
    s \\ \hline 
 \end{array}$
  since this does not 
contain $Q_{s-1}$.
}
 $b =s-1$;
 $v=u+h_{1}$: $h_{1}$ is some shift parameter 
and \framebox{$E_{2 n_{2}}$} is a column
\footnote{We assume that 
$\framebox{$E_{20}$}=
\begin{array}{|c|l}\cline{1-1} 
    \overline{b} & _{v} \\ \cline{1-1} 
   \vdots & \\ \cline{1-1} 
    \overline{b} & _{v-k_{2}+1}\\ \cline{1-1} 
 \end{array}
$
 and  
 $\framebox{$E_{2 k_{2}}$}=
\begin{array}{|c|l}\cline{1-1} 
    \overline{b+1} & _{v} \\ \cline{1-1} 
   \vdots & \\ \cline{1-1} 
    \overline{b+1} & _{v-k_{2}+1}\\ \cline{1-1} 
 \end{array}
$.}
 of the form:
\begin{eqnarray}
 \begin{array}{|c|l}\cline{1-1} 
    \overline{b+1}   & _w \\ \cline{1-1} 
    \vdots & \\ \cline{1-1} 
    \overline{b+1} & _{w-n_{2}+1}\\ \cline{1-1} 
    \overline{b} & _{w-n_{2}} \\ \cline{1-1} 
   \vdots & \\ \cline{1-1} 
    \overline{b} & _{w-k_{2}+1}\\ \cline{1-1} 
 \end{array} 
&=&\frac{Q_{b-1}(w-\frac{2s-b-1}{2}-n_{2})Q_{b}(w-\frac{2s-b}{2}-k_{2})}
      {Q_{b-1}(w-\frac{2s-b-1}{2}-k_{2})Q_{b}(w-\frac{2s-b}{2}-n_{2})}  
\label{tableauxk5} \\ 
&& \times  
 \frac{Q_{b}(w-\frac{2s-b-2}{2})Q_{b+1}(w-\frac{2s-b-1}{2}-n_{2})}
 {Q_{b}(w-\frac{2s-b-2}{2}-n_{2})Q_{b+1}(w-\frac{2s-b-1}{2})}
\nonumber
\end{eqnarray} 
for $2\le b \le s-2$;
\begin{eqnarray}
 \begin{array}{|c|l}\cline{1-1} 
    \overline{s}   & _w \\ \cline{1-1} 
    \vdots & \\ \cline{1-1} 
    \overline{s} & _{w-n_{2}+1}\\ \cline{1-1} 
    \overline{s-1} & _{w-n_{2}} \\ \cline{1-1} 
   \vdots & \\ \cline{1-1} 
    \overline{s-1} & _{w-k_{2}+1}\\ \cline{1-1} 
 \end{array} 
&=&\frac{Q_{s-2}(w-\frac{s}{2}-n_{2})Q_{s-1}(w-\frac{s+1}{2}-k_{2})}
      {Q_{s-2}(w-\frac{s}{2}-k_{2})Q_{s-1}(w-\frac{s+1}{2}-n_{2})} 
\label{tableauxk6} \\ 
&\times&   
 \frac{Q_{s-1}(w-\frac{s-1}{2})
         Q_{s}(w-\frac{s+1}{2}-n_{2})Q_{s}(w-\frac{s-1}{2}-n_{2})}
 {Q_{s-1}(w-\frac{s-1}{2}-n_{2})
      Q_{s}(w-\frac{s-1}{2})Q_{s}(w-\frac{s+1}{2})}
\nonumber
\end{eqnarray}
for $b=s-1$;
$w=u+h_{2}$: $h_{2}$ is some shift parameter; 
$k=k_{1}+k_{2}$. 

For $2\le b \le s-1$, 
\framebox{$E_{1 n_{1}}$} has color $b$ poles at
 $u=-h_{1}+\frac{b-2}{2}+n_{1}+u_{p}^{(b)}$ and 
 $u=-h_{1}+\frac{b-4}{2}+n_{1}+u_{p}^{(b)}$ 
  for $1 \le n_{1} \le k_{1}-1$; at  $u=-h_{1}+\frac{b-2}{2}+u_{p}^{(b)}$
 for $n_{1}=0$ ; at $u=-h_{1}+\frac{b-4}{2}+k_{1}+u_{p}^{(b)}$
 for $n_{1}=k_{1}$. 
The Color $b$ residues at 
$u=-h_{1}+\frac{b-2}{2}+n_{1}+u_{p}^{(b)}$  
 in \framebox{$E_{1 n_{1}}$} and \framebox{$E_{1 \> n_{1}+1}$}
 cancel each other under the BAE (\ref{BAE}). 
 Thus, under the BAE
  (\ref{BAE}), $\sum_{n_{1}=0}^{k_{1}}\framebox{$E_{1 n_{1}}$}$ 
 is free of color $b$ poles
 (see Figure \ref{part-bs}).
\begin{figure}
    \setlength{\unitlength}{1.5pt}
    \begin{center}
    \begin{picture}(270,40) 
     \put(0,0){$
       \framebox{$E_{1 0}$} 
       \stackrel{0}{\longleftarrow }
       \framebox{$E_{1 1}$} 
       \stackrel{1}{\longleftarrow } 
       \cdots 
       \stackrel{n-2}{\longleftarrow }
       \framebox{$E_{1 n-1}$} 
       \stackrel{n-1}{\longleftarrow }
       \framebox{$E_{1 n}$} 
       \stackrel{n}{\longleftarrow } 
       \framebox{$E_{1 n+1}$}
       \stackrel{n+1}{\longleftarrow } 
       \cdots 
       \stackrel{k_{1}-1}{\longleftarrow }
       \framebox{$E_{1 k_{1}}$}
     $}  
    \end{picture}
  \end{center}
  \caption{Partial Bethe-strap structure of  
 $ E_{1 n}$ 
    for color $b$ poles:  
 The number $n$ on the arrow denotes the common color $b$ pole 
 $-h_{1}+\frac{b-2}{2}+n+u_{k}^{(b)}$ 
 of the pair of the tableaux connected by the arrow.   
 This common pole vanishes under the BAE (\ref{BAE}).}
  \label{part-bs}
\end{figure}
 
 \framebox{$E_{2 n_{2}}$} has color $b$ poles at
 $u=-h_{2}+\frac{2s-b}{2}+n_{2}+u_{p}^{(b)}$ and 
 $u=-h_{2}+\frac{2s-b-2}{2}+n_{2}+u_{p}^{(b)}$ 
  for $1 \le n_{2} \le k_{2}-1$; at  
  $u=-h_{2}+\frac{2s-b}{2}+u_{p}^{(b)}$ 
for $n_{2}=0$ ; at 
$u=-h_{2}+\frac{2s-b-2}{2}+k_{2}+u_{p}^{(b)}$
 for $n_{2}=k_{2}$. 
 The color $b$ residues at 
$u=-h_{2}+\frac{2s-b}{2}+n_{2}+u_{p}^{(b)}$
 in \framebox{$E_{2 n_{2}}$} and \framebox{$E_{2, n_{2}+1}$}
 cancel each other under the BAE (\ref{BAE}). 
 Thus, under the BAE
  (\ref{BAE}), $\sum_{n_{2}=0}^{k_{2}}\framebox{$E_{2, n_{2}}$}$ 
 is free of color $b$ poles, 
  so is $S_{k}$. \\ 
$\bullet $ 
The case $b=s$: $S_{k} (k\ge 2)$ is a  
summation of the tableaux (with sign) of the form 
\begin{eqnarray}
g(l,n)&=&
 \begin{array}{|c|l}\cline{1-1} 
    \xi &  \\ \cline{1-1}
    s   & _v \\ \cline{1-1} 
    \vdots & \\ \cline{1-1} 
    s   & _{v-l+1}\\ \cline{1-1} 
    \overline{s} & _{v-l} \\ \cline{1-1}
    s   & _{v-l-1} \\ \cline{1-1} 
   \vdots & \\ \cline{1-1} 
   \overline{s} & _{v-l-2n+2} \\ \cline{1-1}
    s   & _{v-l-2n+1} \\ \cline{1-1}
    \overline{s} & _{v-l-2n} \\ \cline{1-1} 
    \vdots & \\ \cline{1-1} 
    \overline{s} & _{v-k+1} \\ \cline{1-1}
    \zeta &  \\ \cline{1-1}
 \end{array} 
\label{tableauxk7} \\ 
&=&\frac{Q_{s-1}(v-\frac{s-3}{2}-l)Q_{s-1}(v-\frac{s-1}{2}-l-2n)
            }
      {Q_{s-1}(v-\frac{s-3}{2})Q_{s-1}(v-\frac{s-1}{2}-k)
            } 
\nonumber \\ 
&& \times   
 \frac{Q_{s}(v-\frac{s-5}{2})Q_{s}(v-\frac{s-3}{2})
       Q_{s}(v-\frac{s-1}{2}-k)}
      {Q_{s}(v-\frac{s-5}{2}-l)Q_{s}(v-\frac{s-1}{2}-l)
       Q_{s}(v-\frac{s-3}{2}-l-2n)}
             \nonumber \\ 
&& \times  
 \frac{Q_{s}(v-\frac{s+1}{2}-k)}{Q_{s}(v-\frac{s+1}{2}-l-2n)}
  \times \framebox{$\xi $}
       \times \framebox{$\zeta $}, 
\nonumber
\end{eqnarray}
where \framebox{$\xi$} and 
\framebox{$\zeta$} are columns with total 
length $a-k$, which do not contain \framebox{$s$} and 
\framebox{$\overline{s}$};
$v=u+h_{1}$: $h_{1}$ is some shift parameter.
\begin{table}
\begin{center}
 \begin{tabular}{|l|l|l|l|l|} \hline 
\hspace{-6pt} Poles $\setminus$ Zeros & 
 $\frac{s-5}{2}$ & $\frac{s-3}{2}$ & $\frac{s-1}{2}+k$ & 
 $\frac{s+1}{2}+k$          \\ \hline 
\hspace{-6pt} $\frac{s-5}{2}+l$ &(A) 
$l=0$ &(B) $l=1$ & none & none \\ \hline 
\hspace{-6pt} $\frac{s-1}{2}+l$ & 
  none & none &(C) $l=k,n=0$ & none \\ \hline 
\hspace{-6pt} $\frac{s-3}{2}+l+2n$ & 
  none &\hspace{-7pt} (D) $l=n=0$ & none & none \\ \hline 
\hspace{-6pt} $\frac{s+1}{2}+l+2n$ & 
  none & none \hspace{-7pt} & \hspace{-7pt} (E) $l+2n=k-1$ &
   \hspace{-7pt} (F) $l+2n=k$ \hspace{-7pt} \\ \hline 
 \end{tabular} 
\end{center}
\caption{The conditions for occurrence of cancellation of poles and 
zeros in $g(l,n)$ (\ref{tableauxk7}): $g(l,n)$ has a pole at 
$u=y-h_{1}+u_{p}^{(s)}$ and a zero at 
$u=x-h_{1}+u_{p}^{(s)}$, where $y$ is a number listed in the 
first column and $x$ is a number listed in the 
first row. The content of the table indicates when  
these pole and zero cancel each other, i.e. the condition for 
occurrence of  
$x=y$ for $0\le n \le [\frac{k}{2}]$, $0 \le l \le k-2n$.}
\label{can} 
\end{table}
$g(l,n)$ has color $s$ poles, some of which are 
canceled by zeros in the numerator (see Table \ref{can}),
 and remaining 
poles are canceled by 
 the functions \symbol{96}around'  $g(l,n)$ under the BAE 
 (\ref{BAE}) 
 (see Figure \ref{can-odd} and Figure \ref{can-even}). 
 Hereafter we consider the case $k\ge 3$. The case $k=2$ 
 can be treated similarly. \\ 
Denote the sets of $l,n\in {\bf Z}$ 
($0\le n \le [\frac{k}{2}]$, $0 \le l \le k-2n$) 
 satisfying the conditions
 in  Table  \ref{can} as alphabet. Set 
 $G=\{l,n\in {\bf Z} |n=0,1 \le l \le k-1 \}$. \\ 
(1) For $\bar{A} \cap \bar{B} \cap \bar{C} \cap \bar{D} \cap 
\bar{E} \cap  \bar{F} \cap \bar{G}= 
   \{l,n\in {\bf Z} |
   2\le l \le k-2n-2$, $1 \le n \le [\frac{k-1}{2}]-1 \}$, 
   $g(l,n)$ has color $s$ poles
\footnote{
We  assume that these poles at 
$u=-h_{1}+\frac{s-5}{2}+l+u_{i}^{(s)}$,
 $u=-h_{1}+\frac{s+1}{2}+l+2n+u_{j}^{(s)}$,
 $u=-h_{1}+\frac{s-1}{2}+l+u_{p}^{(s)}$
  and 
 $u=-h_{1}+\frac{s-3}{2}+l+2n+u_{q}^{(s)}$ 
 are not coincide each other 
 for any $i,j,p,q \in \{1,2,\dots , N_{s}\} $. 
 If some of these poles coincide, we will 
 need separate consideration. 
 For example, if poles at
  $u=-h_{1}+\frac{s+1}{2}+l+2n+u_{j}^{(s)}$ and 
 $u=-h_{1}+\frac{s-1}{2}+l+u_{p}^{(s)}$ coincide, 
 i.e. $u_{p}^{(s)}=u_{j}^{(s)}+2n+1$, 
 we have to consider not only \symbol{96}nearest functions' 
 $g(l,n+1)$ and $g(l+2,n-1)$ but also \symbol{96}next nearest 
 function' $g(l+2,n)$: 
 $g(l,n)+g(l,n+1)+g(l+2,n-1)+g(l+2,n)$ is free of 
 color $s$ pole at $u=-h_{1}+\frac{s+1}{2}+l+2n+u_{j}^{(s)}=
 -h_{1}+\frac{s-1}{2}+l+u_{p}^{(s)}$ 
 under the BAE (\ref{BAE}). 
 Other cases (including (2)-(12)) 
 can be treated by a similar idea.
 }
    at  
 $u=-h_{1}+\frac{s-5}{2}+l+u_{p}^{(s)}$,
 $u=-h_{1}+\frac{s+1}{2}+l+2n+u_{p}^{(s)}$,
 $u=-h_{1}+\frac{s-1}{2}+l+u_{p}^{(s)}$
  and 
 $u=-h_{1}+\frac{s-3}{2}+l+2n+u_{p}^{(s)}$, 
  and these poles are    
 canceled by $g(l-2,n+1)$, $g(l,n+1)$, $g(l+2,n-1)$ and $g(l,n-1)$ 
 respectively under the BAE (\ref{BAE}). \\ 
(2) For $A \cap D =D=\{l,n\in {\bf Z} | l=n=0\}$, 
$g(l,n)$ has color $s$ poles at 
 $u=-h_{1}+\frac{s+1}{2}+u_{p}^{(s)}$ and 
 $u=-h_{1}+\frac{s-1}{2}+u_{p}^{(s)}$, 
  and these poles are 
  canceled by  $g(0,1)$ and $g(1,0)$ 
 respectively under the BAE (\ref{BAE}). \\ 
(3) For $B\cap G=\{l,n\in {\bf Z} | l=1,n=0\}$, 
$g(l,n)$ has color $s$ poles at 
 $u=-h_{1}+\frac{s+3}{2}+u_{p}^{(s)}$,
 $u=-h_{1}+\frac{s+1}{2}+u_{p}^{(s)}$
  and 
 $u=-h_{1}+\frac{s-1}{2}+u_{p}^{(s)}$, 
  and these poles are    
 canceled by $g(1,1)$, $g(2,0)$ and $g(0,0)$ 
 respectively under the BAE (\ref{BAE}). \\ 
(4) For $\bar{B}\cap G \cap \bar{E} 
  =\{l,n\in {\bf Z} |2\le l \le k-2,n=0\}$, 
  $g(l,n)$ has color $s$ poles at 
 $u=-h_{1}+\frac{s-5}{2}+l+u_{p}^{(s)}$,
 $u=-h_{1}+\frac{s+1}{2}+l+u_{p}^{(s)}$,
 $u=-h_{1}+\frac{s-1}{2}+l+u_{p}^{(s)}$
  and 
 $u=-h_{1}+\frac{s-3}{2}+l+u_{p}^{(s)}$, 
  and these poles are    
 canceled by $g(l-2,1)$, $g(l,1)$, $g(l+1,0)$ and $g(l-1,0)$ 
 respectively under the BAE (\ref{BAE}). \\ 
(5) For $ G \cap E 
  =\{l,n\in {\bf Z} | l=k-1,n=0\}$, 
  $g(l,n)$ has color $s$ poles at 
 $u=-h_{1}+\frac{s-7}{2}+k+u_{p}^{(s)}$,
 $u=-h_{1}+\frac{s-3}{2}+k+u_{p}^{(s)}$
  and 
 $u=-h_{1}+\frac{s-5}{2}+k+u_{p}^{(s)}$, 
  and these poles are    
 canceled by $g(k-3,1)$, $g(k,0)$ and $g(k-2,0)$ 
 respectively under the BAE (\ref{BAE}). \\ 
(6) For $(A\cap F\cap \{k: \ {\rm even}\})\cup 
(A\cap E\cap \{k: \ {\rm odd}\}) 
    =\{l,n\in {\bf Z} | l=0, n=[\frac{k}{2}]\}$, 
    $g(l,n)$ has color $s$ poles at 
 $u=-h_{1}+\frac{s-1}{2}+u_{p}^{(s)}$
  and 
 $u=-h_{1}+\frac{s-3}{2}+2[\frac{k}{2}]+u_{p}^{(s)}$, 
  and these poles are    
 canceled by $g(2,[\frac{k}{2}]-1)$ and $g(0,[\frac{k}{2}]-1)$ 
 respectively under the BAE (\ref{BAE}). \\ 
(7) For $(A \cap \bar{D} \cap \bar{F} \cap \{k: \ {\rm even} \}) 
\cup (A\cap \bar{D} \cap \bar{E} \cap \{k: \ {\rm odd} \}) 
    =\{l,n \in {\bf Z} | l=0,1 \le n \le [\frac{k}{2}]-1 \}$, 
    $g(l,n)$ has color $s$ poles at 
 $u=-h_{1}+\frac{s+1}{2}+2n+u_{p}^{(s)}$,
 $u=-h_{1}+\frac{s-1}{2}+u_{p}^{(s)}$
  and 
 $u=-h_{1}+\frac{s-3}{2}+2n+u_{p}^{(s)}$, 
  and these poles are    
 canceled by $g(0,n+1)$, $g(2,n-1)$ and $g(0,n-1)$ 
 respectively under the BAE (\ref{BAE}). \\
(8) For $(B \cap E \cap \{k: \ {\rm even} \}) \cup 
    (B\cap F \cap \{k: \ {\rm odd} \}) 
    =\{l,n \in {\bf Z} | l=1, n=[\frac{k-1}{2}] \}$, 
    $g(l,n)$ has color $s$ poles at 
 $u=-h_{1}+\frac{s+1}{2}+u_{p}^{(s)}$
  and 
 $u=-h_{1}+\frac{s-1}{2}+2[\frac{k-1}{2}]+u_{p}^{(s)}$, 
  and these poles are    
 canceled by $g(3,[\frac{k-1}{2}]-1)$ and $g(1,[\frac{k-1}{2}]-1)$ 
 respectively under the BAE (\ref{BAE}). \\
(9) For $(B \cap \bar{E} \cap \bar{G} \cap \{k: \ {\rm even} \}) \cup 
    (B\cap \bar{F} \cap \bar{G} \cap \{k: \ {\rm odd} \}) 
    =\{l,n \in {\bf Z} | l=1,1 \le n \le [\frac{k-1}{2}]-1 \}$, 
    $g(l,n)$ has color $s$ poles at 
 $u=-h_{1}+\frac{s+3}{2}+2n+u_{p}^{(s)}$,
 $u=-h_{1}+\frac{s+1}{2}+u_{p}^{(s)}$
  and 
 $u=-h_{1}+\frac{s-1}{2}+2n+u_{p}^{(s)}$, 
  and these poles are    
 canceled by $g(1,n+1)$, $g(3,n-1)$ and $g(1,n-1)$ 
 respectively under the BAE (\ref{BAE}). \\ 
(10) For $(\bar{A} \cap \bar{C} \cap F \cap \{k: \ {\rm even} \}) \cup 
      (\bar{B} \cap \bar{C} \cap F \cap \{k: \ {\rm odd} \}) 
    =\{l,n \in {\bf Z} |l=-2n+k, 1 \le n \le [\frac{k}{2}]-1 \}$, 
    $g(l,n)$ has color $s$ poles at 
 $u=-h_{1}+\frac{s-5}{2}+k-2n+u_{p}^{(s)}$,
 $u=-h_{1}+\frac{s-1}{2}+k-2n+u_{p}^{(s)}$
  and 
 $u=-h_{1}+\frac{s-3}{2}+k+u_{p}^{(s)}$, 
  and these poles are    
 canceled by $g(k-2n-2,n+1)$,
 $g(k-2n+2,n-1)$ and $g(k-2n,n-1)$ 
 respectively under the BAE (\ref{BAE}). \\
(11) For 
$(\bar{B} \cap E \cap \bar{G} \cap \{k :\ {\rm even} \}) \cup 
      (\bar{A} \cap E \cap \bar{G} \cap \{k :\ {\rm odd} \}) 
    =\{l,n \in {\bf Z} |l=-2n+k-1, 1 \le n \le [\frac{k-1}{2}]-1 \}$, 
    $g(l,n)$ has color $s$ poles at 
 $u=-h_{1}+\frac{s-7}{2}+k-2n+u_{p}^{(s)}$,
 $u=-h_{1}+\frac{s-3}{2}+k-2n+u_{p}^{(s)}$
  and 
 $u=-h_{1}+\frac{s-5}{2}+k+u_{p}^{(s)}$, 
  and these poles are    
 canceled by $g(k-2n-3,n+1)$,
 $g(k-2n+1,n-1)$ and $g(k-2n-1,n-1)$ 
 respectively under the BAE (\ref{BAE}). \\
(12) For $ C\cap F  =C 
    =\{l,n \in {\bf Z} |l=k,n=0 \}$, 
    $g(l,n)$ has color $s$ poles at 
   $u=-h_{1}+\frac{s-5}{2}+k+u_{p}^{(s)}$ and 
   $u=-h_{1}+\frac{s-3}{2}+k+u_{p}^{(s)}$, 
   and these poles are    
 canceled by $g(k-2,1)$ and $g(k-1,0)$ 
 respectively under the BAE (\ref{BAE}).
     \\ 
 Thus, under the BAE
  (\ref{BAE}), 
  $\sum_{n=0}^{[\frac{k}{2}]}\sum_{l=0}^{k-2n}g(l,n)$ 
 is free of color $s$ poles,
 so is $S_{k}$.  
\rule{5pt}{10pt} \\ 
\begin{figure}
    \setlength{\unitlength}{1.5pt}
    \begin{center}
    \begin{picture}(120,210) 
     \put(-2,-2){$\oplus $}
     \put(-2,28){$\ominus $} 
     \put(-2,58){$\otimes $}
     \put(-2,88){$\otimes $}
     \put(-2,118){$\otimes $}
     \put(-2,148){$\otimes $}
     \put(-2,178){$\oslash $}
     \put(-2,208){$\diamond $} 
     \put(28,-2){$\triangleleft $}
     \put(28,28){$\star $}
     \put(28,58){$\triangleright $}
     \put(28,88){$\triangleright $}
     \put(28,118){$\bullet $}
     \put(28,148){$\circ $} 
     \put(58,-2){$\triangleleft $}
     \put(58,28){$\star $}
     \put(58,58){$\bullet $}
     \put(58,88){$\circ $}
     \put(88,-2){$\odot $}
     \put(88,28){$\ast $}
     \put(0,25){\vector(0,-1){20}}
     \put(0,55){\vector(0,-1){20}}
     \put(0,85){\vector(0,-1){20}}
     \put(0,115){\vector(0,-1){20}}
     \put(0,145){\vector(0,-1){20}}
     \put(0,175){\vector(0,-1){20}}
     \put(0,205){\vector(0,-1){20}}
     \put(2,56){\vector(1,-2){26}}
     \put(2,86){\vector(1,-2){26}}
     \put(2,116){\vector(1,-2){26}}
     \put(2,146){\vector(1,-2){26}}
     \put(2,176){\vector(1,-2){26}}
     \put(2,206){\vector(1,-2){26}}
     \put(32,56){\vector(1,-2){26}}
     \put(32,86){\vector(1,-2){26}}
     \put(32,116){\vector(1,-2){26}}
     \put(32,146){\vector(1,-2){26}}
     \put(62,56){\vector(1,-2){26}}
     \put(62,86){\vector(1,-2){26}}
     \put(25,0){\vector(-1,0){20}}
     \put(55,0){\vector(-1,0){20}}
     \put(85,0){\vector(-1,0){20}}
     \put(25,30){\line(-1,0){8}}
     \put(13,30){\vector(-1,0){8}}
     \put(55,30){\line(-1,0){8}}
     \put(43,30){\vector(-1,0){8}}
     \put(85,30){\line(-1,0){8}}
     \put(73,30){\vector(-1,0){8}}
     \put(25,60){\line(-1,0){8}}
     \put(13,60){\vector(-1,0){8}}
     \put(55,60){\line(-1,0){8}}
     \put(43,60){\vector(-1,0){8}}
     \put(25,90){\line(-1,0){8}}
     \put(13,90){\vector(-1,0){8}}
     \put(55,90){\line(-1,0){8}}
     \put(43,90){\vector(-1,0){8}}
     \put(25,120){\line(-1,0){8}}
     \put(13,120){\vector(-1,0){8}}
     \put(25,150){\line(-1,0){8}}
     \put(13,150){\vector(-1,0){8}}
    \end{picture}
  \end{center}
  \caption{Partial Bethe-strap structure of  
 $g(l,n)$ for color $s$ poles ($k=7$: odd case):  
 Each term of $g(l,n)$ is marked as some symbol: 
 (1) $\triangleright $, (2) $\oplus $, (3) $\ominus $, (4) $\otimes $, 
 (5) $\oslash $, (6) $\odot $, (7) $\triangleleft $, (8) $\ast $, 
 (9) $\star $, (10) $\circ $, (11) $\bullet $, (12) $\diamond$. 
  The terms connected by the arrow have a common pole, and  
  this common pole vanishes under the BAE (\ref{BAE}). }
  \label{can-odd}
\end{figure}
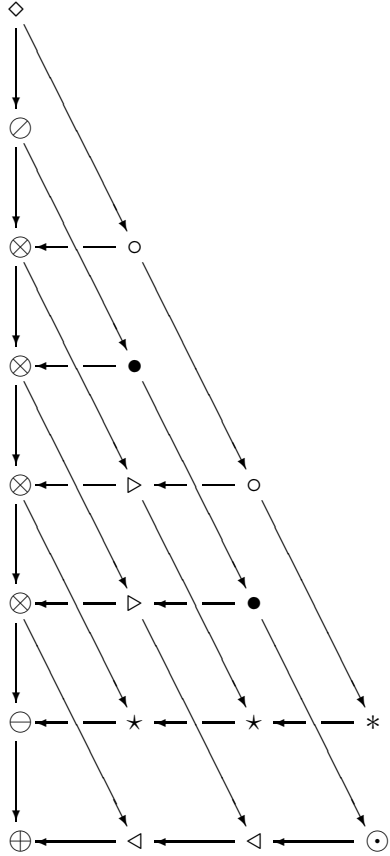
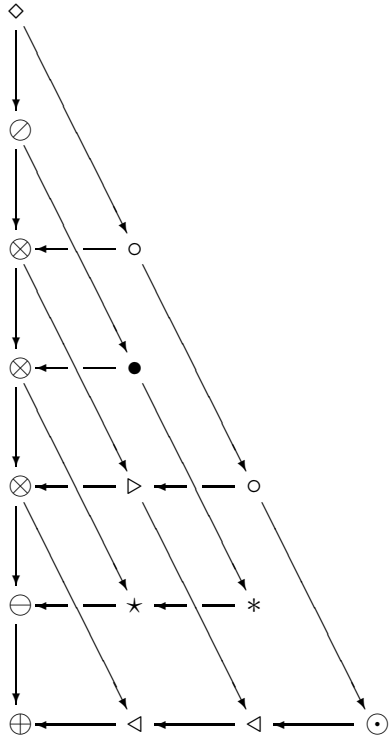
\begin{figure}
    \setlength{\unitlength}{1.5pt}
    \begin{center}
    \begin{picture}(120,210) 
     \put(-2,-2){$\oplus $}
     \put(-2,28){$\ominus $} 
     \put(-2,58){$\otimes $}
     \put(-2,88){$\otimes $}
     \put(-2,118){$\otimes $}
     \put(-2,148){$\oslash $}
     \put(-2,178){$\diamond $} 
     \put(28,-2){$\triangleleft $}
     \put(28,28){$\star $}
     \put(28,58){$\triangleright $}
     \put(28,88){$\bullet $}
     \put(28,118){$\circ $}
     \put(58,-2){$\triangleleft $}
     \put(58,28){$\ast $}
     \put(58,58){$\circ $}
     \put(88,-2){$\odot $}
     \put(0,25){\vector(0,-1){20}}
     \put(0,55){\vector(0,-1){20}}
     \put(0,85){\vector(0,-1){20}}
     \put(0,115){\vector(0,-1){20}}
     \put(0,145){\vector(0,-1){20}}
     \put(0,175){\vector(0,-1){20}}
     \put(2,56){\vector(1,-2){26}}
     \put(2,86){\vector(1,-2){26}}
     \put(2,116){\vector(1,-2){26}}
     \put(2,146){\vector(1,-2){26}}
     \put(2,176){\vector(1,-2){26}}
     \put(32,56){\vector(1,-2){26}}
     \put(32,86){\vector(1,-2){26}}
     \put(32,116){\vector(1,-2){26}}
     \put(62,56){\vector(1,-2){26}}
     \put(25,0){\vector(-1,0){20}}
     \put(55,0){\vector(-1,0){20}}
     \put(85,0){\vector(-1,0){20}}
     \put(25,30){\line(-1,0){8}}
     \put(13,30){\vector(-1,0){8}}
     \put(55,30){\line(-1,0){8}}
     \put(43,30){\vector(-1,0){8}}
     \put(25,60){\line(-1,0){8}}
     \put(13,60){\vector(-1,0){8}}
     \put(55,60){\line(-1,0){8}}
     \put(43,60){\vector(-1,0){8}}
     \put(25,90){\line(-1,0){8}}
     \put(13,90){\vector(-1,0){8}}
     \put(25,120){\line(-1,0){8}}
     \put(13,120){\vector(-1,0){8}}
    \end{picture}
  \end{center}
  \caption{Partial Bethe-strap structure of  
 $g(l,n)$ for color $s$ poles ($k=6$: even case):  
 Each term of $g(l,n)$ is marked as some symbol: 
 (1) $\triangleright $, (2) $\oplus $, (3) $\ominus $, (4) $\otimes $, 
 (5) $\oslash $, (6) $\odot $, (7) $\triangleleft $, (8) $\ast $, 
 (9) $\star $, (10) $\circ $, (11) $\bullet $, (12) $\diamond$. 
 The terms connected by the arrow have a common pole, and  
  this common pole vanishes under the BAE (\ref{BAE}).}
  \label{can-even}
\end{figure}
\section{An extension of the DVF $T_{m}^{(1)}(u)$}
In the previous section, we have given the DVF $T_{m}^{(1)}(u)$ 
(\ref{yoko}) 
only for $m \in {\bf Z} (1\le m \le s-1$). 
In this section,
 we shall extend  $T_{m}^{(1)}(u)$ for  $m \in {\bf C}$. 
 Hereafter, we will consider the case where the quantum space 
 is formally trivial. 
  In this case, the vacuum part of the function 
$\framebox{$a$}_{u}$ (\ref{z+}) is constantly $1$.

 We assume  $T_{m}^{(1)}(u)$ for $m \ge s-1$ ($m \in {\bf Z}$) 
is given by using a deformation of $T_{s-1}^{(1)}(u)$: 
\begin{eqnarray}
 {\cal T}_{m}(u)&=&
\frac{Q_{1}(u-\frac{m}{2})}{Q_{1}(u+\frac{m}{2}-s+1)}
\times
T_{s-1}^{(1)}(u+\frac{m-s+1}{2}).\label{defo}
\end{eqnarray}
For $m=s,s+1,\dots,2s-2 $,
\begin{eqnarray}
 T_{m}^{(1)}(u)&=& {\cal T}_{m}(u)-
  T_{2s-2-m}^{(1)}(u)  
\end{eqnarray}
and for $m=2s-1,2s,\dots $,
\begin{eqnarray}
 T_{m}^{(1)}(u)&=& {\cal T}_{m}(u).  
\end{eqnarray}
This deformation is compatible with the top term 
hypothesis \cite{KS1,KOS}. 
In fact the top term of  $T_{m}^{(1)}(u)$ will be  
$\frac{Q_{1}(u-\frac{m}{2})}{Q_{1}(u+\frac{m}{2})}$, 
which carries $C(s)$ weight $m \omega_{1}$. 
We have confirmed, for several cases, the fact that this top term 
generates $T_{m}^{(1)}(u)$ (cf. Figure \ref{best1}, Figure \ref{best3}).
Moreover, we have also checked, for several cases, 
 that the number of the term in  $T_{m}^{(1)}(u)$
 coincides with ${\rm dim} V(m\omega_{1})$ 
 (cf. Table \ref{dim-c3}, Table \ref{numt1-3}).  
Furthermore $T_{m}^{(1)}(u)$ is free of pole under the 
BAE (\ref{BAE}) ( see  later). We remark that the right hand 
side of (\ref{defo}) is given as a summation of the 
tableaux of the form
\begin{equation}
(-1)^{\sum_{k=1}^{s-1}p(i_{k})}
 \begin{array}{|c|c|c|c|c|c|}
   \multicolumn{6}{c}{} \\ 
   \hline 
   1 & \cdots & 1 & i_{1} & \cdots & i_{s-1}
   \\ \hline 
   \multicolumn{3}{c}{^{\underbrace{\qquad \quad }_{m-s+1}}} &
   \multicolumn{3}{c}{^{\underbrace{\qquad \qquad }_{s-1}}} 
 \end{array}, \label{tb-re} 
\end{equation}
where $\{i_{k}\} \in B((s-1)^{1})$;
 the spectral parameter $u$ is shifted as 
$u-\frac{m-1}{2},u-\frac{m-3}{2},\dots, u+\frac{m-1}{2}$ 
from the left to the right.   
In all these tableaux, the left side is always occupied by the same 
number $1$. This circumstance resembles the $sl(1|s-1)$ case \cite{T3}. 

For $m=s-1,s,\dots,2s-2$, 
every term in   $T_{2s-m-2}^{(1)}(u)$ coincides 
\footnote{We may trace this fact  back to the observation 
(cf \cite{KS1}) that 
the DVF has the following form 
\begin{eqnarray}
 \sum \frac{Q_{a_{1}}(u+\xi_{1})\cdots Q_{a_{k}}(u+\xi_{k})}
           {Q_{a_{1}}(u+\eta_{1})\cdots Q_{a_{k}}(u+\eta_{k})}
           \label{dvf-foot}
\end{eqnarray}
if one neglect the signs originated from 
the grading (\ref{grading}) 
since the transfer matrix is defined as a super-trace of 
a monodromy matrix.
} 
 with a 
term in ${\cal T}_{m}(u)$. 
 This fact can be verified as follows. 
Noting the relation, 
\begin{equation}
\framebox{$1$}_{u}
 \framebox{$\overline{1}$}_{u+s-1}=1,
\end{equation}
we find 
\begin{eqnarray}
&& 
    (-1)^{\sum_{k=1}^{2s-2-m}p(i_{k})}
 \times 
   \begin{array}{|c|c|c|c|c|c|c|c|c|}
   \multicolumn{9}{c}{} \\ \hline 
   1 & \cdots & 1 & i_{1} & \cdots & i_{2s-2-m} &
      \overline{1} & \cdots & \overline{1} 
   \\ \hline 
  \multicolumn{3}{c}{^{\underbrace{\qquad \quad }_{m-s+1}}} & 
  \multicolumn{3}{c}{^{\underbrace{\qquad \qquad \qquad}_{2s-2-m}}} &
  \multicolumn{3}{c}{^{\underbrace{\qquad \quad }_{m-s+1}}}
 \end{array}
\nonumber \\ 
&&= 
 (-1)^{\sum_{k=1}^{2s-2-m}p(i_{k})}
 \times
\begin{array}{|c|c|c|}
    \multicolumn{3}{c}{} \\ \hline 
   i_{1} & \cdots & i_{2s-2-m}  \\ \hline 
    \multicolumn{3}{c}{^{\underbrace{\qquad \qquad \qquad}_{2s-2-m}}} 
 \end{array},
 \label{eqr1}
\end{eqnarray}
where the spectral parameter $u$ is shifted as 
$u-\frac{m-1}{2},u-\frac{m+1}{2},\dots, u+\frac{m-1}{2}$ 
from the left to the right on the left hand side of (\ref{eqr1}); 
$u-\frac{2s-m-3}{2},u-\frac{2s-m-5}{2},\dots, u+\frac{2s-m-3}{2}$ 
from the left to the right on the right hand side
 of (\ref{eqr1}).
 Apparently, the left hand side of (\ref{eqr1}) coincides with 
 a term in ${\cal T}_{m}(u)$ 
  (see (\ref{tb-re}))
 if $\{ i_{j} \} \in B((2s-m-2)^{1})$. 

In the DVF $T_{m}^{(1)}(u)$, 
we assume $m\in {\bf Z}_{\ge 0}$. 
However, in view of the fact that one can construct finite dimensional 
module whose first Kac-Dynkin label is complex number,
 $T_{m}^{(1)}(u)$ will be also valid for $m \in {\bf C}$ 
by \symbol{96}analytic continuation\symbol{39}. 
Namely, we assume that the DVF whose top term carries
 $C(s)$ weight
 \footnote{Note that this weight corresponds to typical representation.} 
  $c \omega_{1}$ ($c\in {\bf C}$; 
 $c \ne 0,1,\dots, s-2,s,s+1,\dots,2s-2$) 
is given by the right hand side of the (\ref{defo}) 
for $m=c \in {\bf C}$. 
 \footnote{See Appendix A for an example of
  $T_{c}^{(1)}(u)$ for $c\in {\cal C}$.}. 
  Then we find the following 
  theorem, which is a generalization of Theorem \ref{th-yoko}. 
\begin{theorem}\label{main} For any $c \in {\bf C}$, 
the DVF $T_{c}^{(1)}(u)$  
is free of poles under the condition that
the BAE {\rm (\ref{BAE})} is valid.   
\end{theorem}
{\em Proof}. 
Thanks to the Theorem \ref{th-yoko},  
 $T_{s-1}^{(1)}(u+\frac{c-s+1}{2})$ and 
 $T_{2s-2-m}^{(1)}(u)$ ($m\in \{s,s+1,\dots, 2s-2 \}$)
 are free of poles under 
the BAE {\rm (\ref{BAE})}. Then we 
 have only to show that the function ${\cal T}_{c}(u)$, i.e. 
 (\ref{defo}) for  $m=c \in {\bf C}$, 
 is free of pole at $u=u_{k}^{(1)}-\frac{c}{2}+s-1$ 
 ($k=1,\dots ,N_{1}$). 
 We will show that 
 \begin{equation} 
 T_{s-1}^{(1)}(u+\frac{c-s+1}{2})=
 \sum_{\{i_{k}\} \in B((s-1)^{1})}
(-1)^{\sum_{k=1}^{s-1}p(i_{k})}
 \times
\begin{array}{|c|c|c|c|} \hline 
   i_{1} & i_{2} & \cdots & i_{s-1}  \\ \hline  
 \end{array}
 \label{ts-1} 
 \end{equation}
  is divisible 
 by $Q_{1}(u+\frac{c}{2}-s+1)$. 
In the set $\{ \framebox{$a$}_{u+\xi}: a \in J, \xi \in {\bf C} \}$, 
 only $\framebox{$1$}_{u+\frac{c+3}{2}-s}$, 
 $\framebox{$2$}_{u+\frac{c+3}{2}-s}$,
  $\framebox{$\bar{1}$}_{u+\frac{c-1}{2}}$
   and $\framebox{$\bar{2}$}_{u+\frac{c-1}{2}}$ 
 have $Q_{1}(u+\frac{c}{2}-s+1)$ in their numerators. 
 So we have only to show that every term
  in (\ref{ts-1}) contains  
  at least one of them. 
 Then all we have to do is to show that $i_{1}=1$ or $2$, 
  or $i_{s-1}=\bar{1}$ or $\bar{2}$
 in (\ref{ts-1}) 
 since the argument of  $\framebox{$i_{j}$}_{u+\frac{c+1}{2}-s+j}$ 
 ($j=1,2,\dots , s-1$) 
 in (\ref{ts-1}) becomes $u+\frac{c+3}{2}-s$ 
 (resp. $u+\frac{c-1}{2}$) 
  only when it's subscript is $j=1$ (resp.$j=s-1$).  
  
Now we assume $i_{1}\ne 1,2$ and $i_{s-1}\ne \overline{1},\overline{2}$, 
which will lead contradiction.
 From the admissibility conditions (\ref{adm1})-(\ref{adm3}), 
 there is at least one $d \in J$ such that both $d$ and $\bar{d}$ 
 appear in the row since 
 $i_{j} \in \{3,4,\dots,s,\bar{s},\dots,\bar{3}\}$ and 
 the length of the row is $s-1$. If  $d_{min}$ is minimum of such $d$,
  then we find that the tableaux in right hand side of (\ref{ts-1}) 
  have the following form: 
\begin{equation}
  \begin{array}{|c|c|c|c|c|}\hline
    \xi & d_{min} & \eta & \overline{d_{min}} & \zeta  \\
    \hline 
  \end{array},
\end{equation}
where 
  $\xi$ contains only the elements in $\{3,4,\dots,d_{min}-1\}$; 
  $\zeta$ contains only the elements  in 
  $\{\overline{d_{min}-1},\overline{d_{min}-2},\dots,\bar{3}\}$; 
  $\beta$ and $\overline{\beta}$ do not appear simultaneously 
  in $\xi$ and $\zeta$. Then the following inequality is valid:
\begin{equation}
 |\xi|+|\zeta| \le d_{min}-3, 
\end{equation} 
 where $|\xi|$ and $|\zeta|$ are the length of $\xi$ and $\zeta$ 
 respectively. On the other hand, from the admissibility condition 
 (\ref{adm1})-(\ref{adm3}), we have
\begin{equation}
 d_{min} \le s-(|\eta|+1), 
\end{equation} 
 where $|\eta|$ is the  length of $\eta$. 
 These inequalities lead contradiction: 
\begin{equation}
s-1=|\xi|+2+|\eta|+|\zeta| \le (d_{min}-3)+2+(s-d_{min}-1)=s-2. 
\end{equation} 
 In the proof the Theorem \ref{th-yoko},
  we need not make use of the factor 
 $Q_{1}(u+\frac{c}{2}-s+1)$ to prove the fact that   
  $T_{s-1}^{(1)}(u+\frac{c-s+1}{2})$ does not have
   a color $1$ pole under the BAE 
  {\rm (\ref{BAE})}. So division by  $Q_{1}(u+\frac{c}{2}-s+1)$ 
  does not influence the proof of the pole-freeness of  
  $T_{s-1}^{(1)}(u+\frac{c-s+1}{2})$ under the BAE 
  {\rm (\ref{BAE})}. Therefore the function ${\cal T}_{c}(u)$  
  is free of poles under 
the BAE {\rm (\ref{BAE})}, so is $T_{c}^{(1)}(u)$.   
 \rule{5pt}{10pt} \\ 
 To construct a transfer matrix whose eigenvalue formula 
 is given by $T_{c}^{(1)}(u)$, one may be able to use the $R$ 
 matrix which is constructed by tensor product graph method 
\cite{DGLZ2}.    
  
Using the function ${\cal T}_{a-2}(u)$ (\ref{defo}), 
we may define the DVF 
$ T_{1}^{(a)}(u) $ ($\in \{2,3,\dots,s \}$)
whose top term will carry \symbol{96}fundamental weight' $\omega_{a}$ 
\begin{equation}
T_{1}^{(a)}(u)={\cal T}_{a-2}(u)-T_{a-2}^{(1)}(u) \qquad 
 a\in \{ 2,3,\dots s\}. \label{defo2}
\end{equation}
We remark that the right hand 
side of (\ref{defo}) for $m=a-2$ ($a\in \{ 2,3,\dots s\}$) 
is given as a 
 summation of the tableaux of the form 
\begin{equation}
(-1)^{\sum_{k=1}^{s-1}p(i_{k})}
 \times 
 \begin{array}{|c|c|c|c|c|c|}
  \multicolumn{6}{c}{} \\ \hline 
   i_{1} & \cdots & i_{s-1} & \overline{1} & \cdots & \overline{1}
   \\ \hline 
   \multicolumn{3}{c}{^{\underbrace{\qquad \quad }_{s-1}}} &
   \multicolumn{3}{c}{^{\underbrace{\qquad \qquad }_{s+1-a}}} 
 \end{array},
\end{equation}
where $\{i_{k}\} \in B((s-1)^{1})$;
the spectral parameter $u$ is shifted as 
$u-\frac{2s-a-1}{2},u-\frac{2s-a-3}{2},\dots, u+\frac{2s-a-1}{2}$ 
from the left to the right. 
For $a=2,3,\dots,s$, 
every term in $T_{a-2}^{(1)}(u)$ 
(in the right hand side of (\ref{defo2})) coincides with a 
term in ${\cal T}_{a-2}(u)$.
 This fact can be verified as follows. 
Noting the relation, 
\begin{equation}
\framebox{$1$}_{u}
 \framebox{$\overline{1}$}_{u+s-1}=1,
\end{equation}
we find 
\begin{eqnarray}
&& 
(-1)^{\sum_{k=1}^{a-2}p(j_{k})}
 \times 
\begin{array}{|c|c|c|c|c|c|c|c|c|}
    \multicolumn{9}{c}{}  \\ \hline 
   1 & \cdots & 1 & j_{1} & \cdots & j_{a-2} &
      \overline{1} & \cdots & \overline{1} 
   \\ \hline 
  \multicolumn{3}{c}{^{\underbrace{\qquad \quad }_{s+1-a}}} & 
  \multicolumn{3}{c}{^{\underbrace{\qquad \qquad \qquad}_{a-2}}} &
  \multicolumn{3}{c}{^{\underbrace{\qquad \quad }_{s+1-a}}}
 \end{array}
\nonumber \\ 
&&=
 (-1)^{\sum_{k=1}^{a-2}p(j_{k})}
 \times 
\begin{array}{|c|c|c|}
    \multicolumn{3}{c}{} \\ \hline 
   j_{1} & \cdots & j_{a-2}  \\ \hline 
    \multicolumn{3}{c}{^{\underbrace{\qquad \qquad \qquad}_{a-2}}} 
 \end{array},
 \label{eqr3}
\end{eqnarray}
where the spectral parameter $u$ is shifted as 
$u-\frac{2s-a-1}{2},u-\frac{2s-a-3}{2},\dots, u+\frac{2s-a-1}{2}$ 
from the left to the right on the left hand side of (\ref{eqr3}); 
$u-\frac{a-3}{2},u-\frac{a-5}{2},\dots, u+\frac{a-3}{2}$ 
from the left to the right on the tableau in the right hand side
 of (\ref{eqr3}).
 Apparently, the left hand side of (\ref{eqr3}) coincides with 
 a term in ${\cal T}_{a-2}(u)$ if $\{ j_{k} \} \in B((a-2)^{1})$. 

The top term of $T_{1}^{(a)}(u)$ (\ref{defo2}) will be  
$(-1)^{a-1}
 \frac{Q_{a}(u-\frac{1}{t_{a}})}{Q_{a}(u+\frac{1}{t_{a}})}$, 
which carries $C(s)$ 
weight 
$\omega_{a}$. 
In fact $T_{1}^{(a)}(u)$ 
contains a term of the form: 
\begin{eqnarray}
&& 
(-1)^{a-1}
 \times 
\begin{array}{|c|c|c|c|c|c|c|c|c|c|}
    \multicolumn{10}{c}{}          \\ \hline 
   1 & \cdots & 1 & 2 & 3 & \cdots & a &
      \overline{1} & \cdots & \overline{1} 
   \\ \hline 
  \multicolumn{3}{c}{^{\underbrace{\qquad \quad }_{s-a}}} & 
  \multicolumn{4}{c}{^{\underbrace{\qquad \qquad \qquad}_{a-1}}} &
  \multicolumn{3}{c}{^{\underbrace{\qquad \quad }_{s+1-a}}}
 \end{array}
\nonumber \\ 
&&=
 (-1)^{a-1}
 \frac{Q_{a}(u-\frac{1}{t_{a}})}{Q_{a}(u+\frac{1}{t_{a}})},
 \label{eqr4}
\end{eqnarray}
where the spectral parameter $u$ is shifted as 
$u-\frac{2s-a-1}{2},u-\frac{2s-a-3}{2},\dots, u+\frac{2s-a-1}{2}$ 
from the left to the right on the left hand side of (\ref{eqr4}).
 Apparently, the left hand side of (\ref{eqr4}) coincides with 
 a term in ${\cal T}_{a-2}(u)$; dose not formally coincide with 
 any term in $T_{a-2}^{(1)}(u)$. 
Moreover, we have checked, for several cases, 
 that the number of the terms in  $T_{1}^{(a)}(u)$ (\ref{defo2})
 coincides with ${\rm dim}V(\omega_{a})$ 
 (cf. Table \ref{dim-c3}, Table \ref{numt3}). 
 
Owing to the Theorem \ref{main}, we have: 
\begin{theorem} 
For any $a \in \{2,3,\dots,s \}$, 
 $T_{1}^{(a)}(u)$ (\ref{defo2}) 
is free of poles under the condition that
the BAE {\rm (\ref{BAE})} is valid.   
\end{theorem}
\section{On functional relations among DVFs}
Now we briefly mention the functional relations among DVFs.
From the admissibility conditions 
(\ref{adtate1})-(\ref{adtate3}) and 
(\ref{adm1})-(\ref{adm3}),  
the following relation holds. 
\begin{equation}
T_{1}^{(1)}(u-\frac{1}{2})T_{1}^{(1)}(u+\frac{1}{2})
=T_{2}^{(1)}(u) +{\cal T}^{2}(u) .
\end{equation}
This functional relation 
 \footnote{
 This functional relation is not complete one. 
 In fact, we do not know general expressions
  of this type of functional relations for 
  $T_{m}^{(1)}(u)$ ($m \in \{2,3,\dots, 2s-1 \}$) 
  since we are lack of DVFs labelled by Young superdiagrams 
  with shape $(m^{a})$ for $m,a \ge 2$.  
  We left this point for future study. 
 } 
 may be related to 
 specialization of the Hirota bilinear 
difference equation \cite{H}. 
Note however that there is another functional relation, 
which comes from a one parameter family of finite dimensional 
representations.  For instance,
 ${\cal T}_{c}(u)$ satisfies 
\begin{equation}
 {\cal T}_{c}(u-\frac{d}{2}){\cal T}_{c}(u+\frac{d}{2})
 ={\cal T}_{c-d}(u) {\cal T}_{c+d}(u),
\end{equation}
 where $c,d\in {\bf C}$. 
 As special cases, this reduces to 
 the following relations:
\begin{eqnarray}
&& \hspace{-20pt}
(T_{m}^{(1)}(u-\frac{1}{2})+T_{1}^{(m+2)}(u-\frac{1}{2}))
(T_{m}^{(1)}(u+\frac{1}{2})+T_{1}^{(m+2)}(u+\frac{1}{2})) 
\\ 
&& =
(T_{m-1}^{(1)}(u)+T_{1}^{(m+1)}(u))
(T_{m+1}^{(1)}(u)+T_{1}^{(m+3)}(u)) 
\quad m \in \{1,2,\dots, s-3 \}, \nonumber \\ 
&& 
(T_{s-2}^{(1)}(u-\frac{1}{2})+T_{1}^{(s)}(u-\frac{1}{2}))
(T_{s-2}^{(1)}(u+\frac{1}{2})+T_{1}^{(s)}(u+\frac{1}{2})) 
\\ 
&& \hspace{30pt}=
(T_{s-3}^{(1)}(u)+T_{1}^{(s-1)}(u))
T_{s-1}^{(1)}(u), \nonumber \\
&& \hspace{-20pt} 
T_{s-1}^{(1)}(u-\frac{1}{2})T_{s-1}^{(1)}(u+\frac{1}{2})
=
(T_{s-2}^{(1)}(u)+T_{1}^{(s)}(u))
(T_{s}^{(1)}(u)+T_{s-2}^{(1)}(u)) ,  \\ 
&& 
(T_{s}^{(1)}(u-\frac{1}{2})+T_{s-2}^{(1)}(u-\frac{1}{2}))
(T_{s}^{(1)}(u+\frac{1}{2})+T_{s-2}^{(1)}(u+\frac{1}{2})) 
\\ 
&& \hspace{30pt}=
T_{s-1}^{(1)}(u)
(T_{s+1}^{(1)}(u)+T_{s-3}^{(1)}(u)), \nonumber \\
&& \hspace{-20pt}
(T_{m}^{(1)}(u-\frac{1}{2})+T_{2s-m-2}^{(1)}(u-\frac{1}{2}))
(T_{m}^{(1)}(u+\frac{1}{2})+T_{2s-m-2}^{(1)}(u+\frac{1}{2})) 
\nonumber 
\\ 
&& =
(T_{m-1}^{(1)}(u)+T_{2s-m-1}^{(1)}(u))
(T_{m+1}^{(1)}(u)+T_{2s-m-3}^{(1)}(u)) \nonumber \\
&& \hspace{120pt}  m \in \{s+1,s+2,\dots, 2s-3 \}, \\
&& \hspace{-20pt}
(T_{2s-2}^{(1)}(u-\frac{1}{2})+1)
(T_{2s-2}^{(1)}(u+\frac{1}{2})+1) =
T_{2s-1}^{(1)}(u)
(T_{2s-3}^{(1)}(u)+T_{1}^{(1)}(u)) , \\ 
&& 
T_{2s-1}^{(1)}(u-\frac{1}{2})T_{2s-1}^{(1)}(u+\frac{1}{2}) =
(T_{2s-2}^{(1)}(u)+1) T_{2s}^{(1)}(u),  \\ 
&& \hspace{-20pt} 
T_{m}^{(1)}(u-\frac{1}{2}) T_{m}^{(1)}(u+\frac{1}{2})=
T_{m-1}^{(1)}(u) T_{m+1}^{(1)}(u)
\quad m \in \{2s,2s+1,\dots \}. 
\end{eqnarray}
Let $T_{m}^{(a)}(u)$ 
($m \in {\bf Z}_{\ge 0}$, $a \in \{ 2,3,\dots , s \}$) 
be the DVF
\footnote{We assume $T_{0}^{(a)}(u)=1$.}
 whose top term 
\footnote{In this case, the top term is supposed to be 
proportional to 
$\frac{Q_{a}(u-\frac{m}{t_{a}})}{Q_{a}(u+\frac{m}{t_{a}})}$.}
carries $C(s)$ weight  
 $m\omega_{a}$ ($a \in \{ 2,3,\dots , s \}$). 
As for specific values of $(m,a)$, 
we have already given the expression of $T_{m}^{(a)}(u)$. 
In general, we conjecture that  $T_{m}^{(a)}(u)$ is given 
 as a solution of 
the following set of functional relations: 

For $C(3)$; $m \in {\bf Z}_{\ge 1}$, 
\begin{eqnarray}
\hspace{-45pt} &&  
T_{-m}^{(1)}(u-\frac{1}{2}) T_{-m}^{(1)}(u+\frac{1}{2})  = 
        \left\{
          \begin{array}{ll}
            T_{-2}^{(1)}(u)(T_{1}^{(2)}(u)+1)
             & {\rm for} \quad m=1 \\ 
            T_{-m+1}^{(1)}(u) T_{-m-1}^{(1)}(u) 
              & {\rm for} \quad m\in {\bf Z}_{\ge 2}
          \end{array}
        \right. ,  
        \label{t-sys1} \\ 
\hspace{-45pt} &&    T_{2m}^{(2)}(u-\frac{1}{2}) 
T_{2m}^{(2)}(u+\frac{1}{2}) 
  = T_{2m+1}^{(2)}(u) T_{2m-1}^{(2)}(u)\nonumber \\ 
\hspace{-45pt} && \hspace{145pt}  +
        T_{-2m}^{(1)}(u)T_{m}^{(3)}(u-\frac{1}{2}) 
        T_{m}^{(3)}(u+\frac{1}{2}) ,
        \label{t-sysC-5} \\ 
\hspace{-45pt} &&    T_{2m-1}^{(2)}(u-\frac{1}{2})
         T_{2m-1}^{(2)}(u+\frac{1}{2})  = 
        T_{2m}^{(2)}(u) T_{2m-2}^{(2)}(u) \nonumber \\ 
\hspace{-45pt} &&  \hspace{170pt} +
        T_{-2m+1}^{(1)}(u)T_{m-1}^{(3)}(u) T_{m}^{(3)}(u), 
        \label{t-sysC-6} \\ 
\hspace{-45pt} &&    T_{m}^{(3)}(u-1) T_{m}^{(3)}(u+1)  =
        T_{m+1}^{(3)}(u) T_{m-1}^{(3)}(u)+
        T_{2m}^{(2)}(u) .  
        \label{t-sysC-7}
\end{eqnarray}
For $C(s)$; $s\ge 4$; $m\in {\bf Z}_{\ge 1}$, 
\begin{eqnarray}
\hspace{-45pt} &&     T_{-m}^{(1)}(u-\frac{1}{2})
 T_{-m}^{(1)}(u+\frac{1}{2})  = 
\left\{
          \begin{array}{ll}
            T_{-2}^{(1)}(u)(T_{1}^{(2)}(u)+1)
             & {\rm for} \quad m=1 \\ 
            T_{-m+1}^{(1)}(u) T_{-m-1}^{(1)}(u) 
              & {\rm for} \quad m\in {\bf Z}_{\ge 2}
          \end{array}
        \right.
         , \label{t-sys1-2} \\ 
\hspace{-45pt} &&     T_{m}^{(2)}(u-\frac{1}{2}) 
T_{m}^{(2)}(u+\frac{1}{2})  = 
        T_{m+1}^{(2)}(u) T_{m-1}^{(2)}(u)+
        T_{-m}^{(1)}(u) T_{m}^{(3)}(u), \\
\hspace{-45pt} &&     T_{m}^{(a)}(u-\frac{1}{2}) 
T_{m}^{(a)}(u+\frac{1}{2})  = 
        T_{m+1}^{(a)}(u) T_{m-1}^{(a)}(u)+
        T_{m}^{(a-1)}(u) T_{m}^{(a+1)}(u) \nonumber \\ 
\hspace{-45pt} && \hspace{220pt} 3 \le a \le s-2, 
        \label{2t-sysC-4} \\
\hspace{-45pt} &&    T_{2m}^{(s-1)}(u-\frac{1}{2})
                     T_{2m}^{(s-1)}(u+\frac{1}{2}) 
  = T_{2m+1}^{(s-1)}(u) T_{2m-1}^{(s-1)}(u)\nonumber \\ 
\hspace{-45pt} && \hspace{145pt}  +
        T_{2m}^{(s-2)}(u)T_{m}^{(s)}(u-\frac{1}{2}) 
         T_{m}^{(s)}(u+\frac{1}{2}) ,
        \label{2t-sysC-5} \\ 
\hspace{-45pt} &&    T_{2m-1}^{(s-1)}(u-\frac{1}{2}) 
                   T_{2m-1}^{(s-1)}(u+\frac{1}{2})  = 
        T_{2m}^{(s-1)}(u) T_{2m-2}^{(s-1)}(u) \nonumber \\ 
\hspace{-45pt} &&  \hspace{170pt} +
        T_{2m-1}^{(s-2)}(u)T_{m-1}^{(s)}(u) T_{m}^{(s)}(u), 
        \label{2t-sysC-6} \\ 
\hspace{-45pt} &&    T_{m}^{(s)}(u-1) T_{m}^{(s)}(u+1)  =
        T_{m+1}^{(s)}(u) T_{m-1}^{(s)}(u)+
        T_{2m}^{(s-1)}(u) .  
        \label{t-sys7}
\end{eqnarray}
{\em Remark}: 
Apparently the solutions of these functional relations 
are not polynomials of  
$T_{-1}^{(1)},T_{1}^{(2)},\dots,T_{1}^{(s)}$ but 
rational functions. Nevertheless we have confirmed, 
for several cases,  
the fact that the solutions have the form (\ref{dvf-foot})  
if we express $T_{-1}^{(1)},T_{1}^{(2)},\dots,T_{1}^{(s)}$  
 by using the $Q_{a}$-functions. 
 We also note that these functional relations 
 have determinant or pfaffian solutions 
 whose matrix elements 
 are 
$T_{-1}^{(1)},T_{1}^{(2)},\dots,T_{1}^{(s)}$ 
if one change the boundary conditions for $m=1$ in 
(\ref{t-sys1}) and (\ref{t-sys1-2}) 
to 
$T_{-1}^{(1)}(u-\frac{1}{2}) 
T_{-1}^{(1)}(u+\frac{1}{2})=T_{-2}^{(1)}(u)$.

As an example, we give the number of the terms in $T_{m}^{(a)}(u)$
 for $C(3)$ for several cases
  (see Table \ref{numt1-3} and Table \ref{numt3}). 
In general, we conjecture that the 
 number of the terms in $T_{m}^{(a)}(u)$ is given as follows:  
\begin{eqnarray}
 {\cal N}_{m}^{(a)}=
 \left\{
  \begin{array}{lll} 
   {\rm dim}V(m\omega_{1}) & {\rm if} & a=1, \\ 
   \sum_{\{ k_{j} \} \in K_{(a,m)}}
   {\rm dim}V(-k_{1}\omega_{1}+k_{2}\omega_{2}+\cdots 
      +k_{a}\omega_{a}) & & \\
     \qquad \qquad \qquad 
     {\rm if} \quad a \in \{ 2,3,\dots s-1 \}, && \\ 
    {\rm dim}V(m\omega_{s}) & {\rm if} & a=s ,
  \end{array} 
 \right.
\end{eqnarray}
where $K_{(a,m)}=\{ \{ k_{j} \}  |
 k_{1}+k_{2}+\cdots +k_{a}
 \le m, k_{j} \equiv m \delta_{j a} \pmod{2} , 
 k_{j} \in {\bf Z}_{\ge 0} \}$. 
 For example, for $C(3)$ case, 
 we have (cf. Tables \ref{dim-c3}, \ref{numt1-3}, \ref{numt3}) 
\begin{table}
\begin{center}
 \begin{tabular}{|c|c|c|c|c|c|} \hline 
 $m$  & 1 & 2  & 3 & 4 & 5   \\ \hline 
 ${\cal N}_{m}^{(2)}$   & $15$ & $65$ & $175$ & $385$ & $735$  \\ \hline 
 ${\cal N}_{m}^{(3)}$   & $10$ & $35$ & $84$ & $165$ & $286$  \\ \hline 
 \end{tabular} 
\end{center}
\caption{The number ${\cal N}_{m}^{(a)}$ of the terms 
 in $T_{m}^{(a)}(u)$ ($a =2,3$)
 for $C(3)$.}
\label{numt3} 
\end{table}
\begin{eqnarray}
{\cal N}_{l}^{(1)}&=&{\rm dim}V(l\omega_{1})\qquad l \in {\bf R}, 
 \nonumber \\ 
{\cal N}_{1}^{(2)}&=&{\rm dim}V(\omega_{2}),  \nonumber \\
{\cal N}_{2}^{(2)}&=&{\rm dim}V(-2\omega_{1})+{\rm dim}V(2\omega_{2}), 
\nonumber \\ 
{\cal N}_{3}^{(2)}&=&{\rm dim}V(-2\omega_{1}+\omega_{2})+
 {\rm dim}V(3\omega_{2}), \nonumber \\ 
{\cal N}_{4}^{(2)}&=&{\rm dim}V(-4\omega_{1})+
 {\rm dim}V(-2\omega_{1}+2\omega_{2})+{\rm dim}V(4\omega_{2}),
 \nonumber \\
{\cal N}_{5}^{(2)}&=&{\rm dim}V(-4\omega_{1}+\omega_{2})+
 {\rm dim}V(-2\omega_{1}+3\omega_{2})+{\rm dim}V(5\omega_{2}),
 \nonumber \\ 
&& {\Large \cdots \cdots},  \nonumber \\
{\cal N}_{m}^{(3)}&=&{\rm dim}V(m\omega_{3}) 
\qquad m \in {\bf Z}_{\ge 0}. 
\end{eqnarray}
These relations seem to suggest decompositions 
of the auxiliary spaces 
 similar to (\ref{decom}). 
\eqreset
\section{Summary and discussion}
In the present paper, we have executed an analytic Bethe ansatz 
 based on the Bethe ansatz equation (\ref{BAE}) with distinguished
 simple root system of the type 1 Lie superalgebra $C(s)$. 
 Eigenvalue formulae of transfer matrices in 
 DVFs are proposed not only for tensor-like representations 
 but also for a one parameter
  family of finite dimensional representations. 
  The key is the top term hypothesis 
   and pole-freeness under the BAE. 
 Pole-freeness of the DVF was proven.
 Functional relations have been discussed for the DVFs. 
 To the author' knowledge, this paper is the first 
 {\em systematic } treatment of the analytic Bethe ansatz related to 
 the Lie superalgebra $C(s)$.
  
  The Lie superalgebras or their
   quantum analogues are not straightforward 
 extension of their non-super counterparts.
 They have several inequivalent sets of simple root systems. 
  In view of this fact, we
   discussed \cite{T2} relations among sets of the Bethe 
 ansatz equations for any simple root systems 
 in relation to the Weyl supergroup for $sl(r+1|s+1)$ case. 
 A similar argument will be also valid for $C(s)$ case.
 
It will be an interesting problem to extend similar 
analysis for the type 2 Lie superalgebras 
$B(r|s)$ and $D(r|s)$. 
In this case, there is no parameter family of finite dimensional 
representations. 
 To construct DVFs related to spnorial representations 
 of the type 2 Lie superalgebras 
 will be rather cumbersome problem. 

 Analytic Bethe ansatz attracts our interest not only in the context of 
 solvable lattice models but also representation theory in mathematics.  
 DVFs in analytic Bethe ansatz may be viewed \cite{KS1,KOS} as 
  characters of representations of Yangians.  
  A remarkable coincidence between 
  currents of deformed Virasoro algebras  and  DVFs was  
  pointed out in Ref. \cite{FR}. 
 We hope that this paper not only provides us practical formulae in 
 statistical physics but also contributes to the development of 
 such interplay between mathematics and  physics. 
\\
{\bf Acknowledgment} \\  
The author would like to thank Prof. A. Kuniba for discussions. 
This work is supported by a Grant-in-Aid for 
JSPS Research   Fellows from the Ministry of Education, Science and 
Culture of Japan.
\eqreset
\renewcommand{\theequation}{A.\arabic{equation}}
\section*{Appendix A An example of the DVF}
In this section
\footnote{We assume that the vacuum part of the DVF is formally trivial.}, 
 we present an example of the DVF
 $T_{c}^{(1)}(u)$ ($c\in {\bf C}$; $c\ne 0,1,3,4$)
 and the Theorem \ref{main} 
for $C(3)$; 
$J_{+}=\{1,\bar{1} \};J_{-}=\{2,3,\bar{3},\bar{2} \}$ case: 
\begin{eqnarray}
T_{c}^{(1)}(u) &=& 
\frac{Q_{1}({\frac{-c}{2}} + u)}{Q_{1}({\frac{c}{2}}-2 + u)} 
T_{2}^{(1)}(u+\frac{c-2}{2}) \nonumber \\ 
&=&
\frac{Q_{1}({\frac{-c}{2}} + u)}{Q_{1}({\frac{c}{2}} + u)} 
  +
\frac{Q_{1}({\frac{-c}{2}} + u)}
     {Q_{1}({\frac{-8+c}{2}} + u)}
  +
  \frac{Q_{1}(\frac{-c}{2} + u)
      Q_{1}(\frac{-4+c}{2}+ u)}{Q_{1}(\frac{-6+c}{2} + u)
      Q_{1}(\frac{-2+c}{2}+ u)} 
       \nonumber \\ &-& 
  \frac{Q_{1}(\frac{-c}{2}+ u)
      Q_{2}(-\frac{9-c}{2} + u)}{
      Q_{1}(\frac{-8+c}{2}+ u)Q_{2}(\frac{-7+c}{2}+ u)}
      \nonumber \\ &-& 
  \frac{Q_{1}(\frac{-c}{2}+ u)
      Q_{1}(\frac{-4+c}{2} + u)Q_{2}(\frac{-7+c}{2}+ u)}{
      Q_{1}(\frac{-6+c}{2} + u)
      Q_{1}(\frac{-2+c}{2} + u)Q_{2}(\frac{-5+c}{2} + u)} 
       \nonumber \\ &-& 
   \frac{Q_{1}(\frac{-c}{2} + u)
      Q_{1}(\frac{-4+c}{2}+ u)Q_{2}(\frac{-1+c}{2}+ u)}{
      Q_{1}(\frac{-6+c}{2}+ u)
      Q_{1}(\frac{-2+c}{2}+ u)Q_{2}(\frac{-3+c}{2}+ u)} \nonumber \\ &+& 
   \frac{Q_{1}(\frac{-c}{2} + u)
      Q_{1}(\frac{-4+c}{2} + u)Q_{2}(\frac{-7+c}{2}+ u)
      Q_{2}(\frac{-1+c}{2}+ u)}{Q_{1}(\frac{-6+c}{2}+ u)
      Q_{1}(\frac{-2+c}{2}+ u)Q_{2}(\frac{-5+c}{2}+ u)
      Q_{2}(\frac{-3+c}{2}+ u)} 
       \nonumber \\ &-& 
   \frac{Q_{1}(\frac{-c}{2}+ u)
      Q_{2}(\frac{1+c}{2}+ u)}{
      Q_{1}(\frac{c}{2} + u)Q_{2}(\frac{-1+c}{2}+ u)}
      +
  \frac{Q_{1}(\frac{-c}{2} + u)
      Q_{3}(\frac{-9+c}{2}+ u)}{Q_{1}(\frac{-6+c}{2}+ u)
      Q_{3}(\frac{-5+c}{2}+ u)}
       \nonumber \\ &-&  
      \frac{
      Q_{1}(\frac{-c}{2} + u)
      Q_{2}(\frac{-5+c}{2}+ u)Q_{3}(\frac{-9+c}{2}+ u)}
      {Q_{1}(\frac{-6+c}{2}+u)
      Q_{2}(\frac{-7+c}{2}+u)Q_{3}(\frac{-5+c}{2}+ u)} 
       \nonumber \\ &-& 
     \frac{Q_{1}(\frac{-c}{2} + u)
      Q_{2}(\frac{-3+c}{2}+ u)Q_{3}(\frac{-7+c}{2}+ u)
      }{Q_{1}(\frac{-2+c}{2}+ u)
      Q_{2}(\frac{-5+c}{2}+ u)Q_{3}(\frac{-3+c}{2}+ u)} 
       \nonumber \\ &+&
        \frac{Q_{1}(\frac{-c}{2} + u)
      Q_{2}(\frac{-1+c}{2}+ u)Q_{3}(\frac{-7+c}{2}+ u)
      }{
      Q_{1}(\frac{-2+c}{2}+ u)Q_{2}(\frac{-5+c}{2}+ u)
      Q_{3}(\frac{-3+c}{2}+ u)} 
       \nonumber \\ &+& 
  \frac{Q_{1}(\frac{-c}{2} + u)
      Q_{2}(\frac{-7+c}{2}+ u)Q_{3}(\frac{-1+c}{2}+ u)
      }{
      Q_{1}(\frac{-6+c}{2}+ u)Q_{2}(\frac{-3+c}{2}+ u)
      Q_{3}(\frac{-5+c}{2} + u)} 
       \nonumber \\ &-& 
  \frac{Q_{1}(\frac{-c}{2} + u)
      Q_{2}(\frac{-5+c}{2}+ u)Q_{3}(\frac{-1+c}{2}+ u)
      }{Q_{1}(\frac{-6+c}{2} + u)
      Q_{2}(\frac{-3+c}{2}+ u)Q_{3}(\frac{-5+c}{2}+ u)
      }
       \nonumber \\ &+&
        \frac{Q_{1}(\frac{-c}{2} + u)
      Q_{3}(\frac{1+c}{2}+ u)}{Q_{1}(\frac{-2+c}{2}+ u)
      Q_{3}(\frac{-3+c}{2}+ u)} 
       \nonumber \\ &-& 
  \frac{Q_{1}(\frac{-c}{2}+ u)
      Q_{2}(\frac{-3+c}{2}+ u)Q_{3}(\frac{1+c}{2}+ u)}
      {Q_{1}(\frac{-2+c}{2}+ u)
      Q_{2}(\frac{-1+c}{2}+ u)Q_{3}(\frac{-3+c}{2}+ u)
      }
,\label{exdvf} 
\end{eqnarray}
where $T_{2}^{(1)}(u)$ is given in (\ref{t21-ex}).
The first term in the right hand side of (\ref{exdvf}) is 
the top term, which is related to the highest weight 
$c \omega_{1} $. 
Although the DVF (\ref{exdvf}) depends on a continuous parameter 
$c$, thanks to Theorem \ref{main}, (\ref{exdvf}) is free of 
pole under the BAE (\ref{BAE2}). 
\eqreset
\renewcommand{\theequation}{B.\arabic{equation}}
\section*{Appendix B $C(2) \simeq sl(1|2)$ case}
We mainly considered the DVFs related to $C(s)$ ($s\ge 3$) 
in the main text. Many of the formulae in the main text 
are also valid for $C(2)$ case 
\footnote{
In this section, we assume that the vacuum part of 
the DVF is formally trivial.}.
\begin{figure}
    \setlength{\unitlength}{1pt}
    \begin{center}
    \begin{picture}(250,50) 
      \put(22.929,12.9289){\line(1,1){14.14214}}
      \put(22.929,27.07107){\line(1,-1){14.14214}}
      \put(30,20){\circle{20}}
      \put(38.8,25){\line(1,0){32.4}}
      \put(38.8,15){\line(1,0){32.4}}
      \put(80,20){\circle{20}}
      \put(40,20){\line(1,1){14.14214}}
      \put(40,20){\line(1,-1){14.14214}}
      \put(17,0){$ \epsilon -\delta_{1} $}
      \put(75,0){$2\delta_{1} $}
      \put(127,17){$\simeq $}
      \put(172.929,12.9289){\line(1,1){14.14214}}
      \put(172.929,27.07107){\line(1,-1){14.14214}}
      \put(180,20){\circle{20}}
      \put(190,20){\line(1,0){30}}
      \put(230,20){\circle{20}}
      \put(167,0){$ \epsilon^{*} -\delta_{1}^{*} $}
      \put(216,0){$\delta_{1}^{*} -\delta_{2}^{*} $}
  \end{picture}
  \end{center}
  \caption{Dynkin diagram for the Lie superalgebra 
  $C(2) \simeq sl(1|2)$ corresponding to the distinguished simple 
  root system: white circle denotes even root; 
   grey (a cross) circle denotes odd root $\alpha$ with 
   $(\alpha|\alpha)=0$.}
  \label{dynkin-c2}
\end{figure}
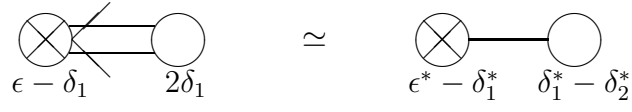
For $C(2)$; $m \in {\bf Z}_{\ge 1}$, the relations 
(\ref{t-sys1})-(\ref{t-sys7}) 
take the following form: 
\begin{eqnarray}
\hspace{-45pt} &&  
T_{-m}^{(1)}(u-\frac{1}{2}) T_{-m}^{(1)}(u+\frac{1}{2})  = 
        \left\{
          \begin{array}{ll}
            T_{-2}^{(1)}(u)(T_{1}^{(2)}(u)+1)
             & {\rm for} \quad m=1, \\ 
            T_{-m+1}^{(1)}(u) T_{-m-1}^{(1)}(u) 
              & {\rm for} \quad m\in {\bf Z}_{\ge 2},
          \end{array}
        \right.   \label{eq-c2-1} \\  
\hspace{-45pt} &&    T_{m}^{(2)}(u-1) T_{m}^{(2)}(u+1)  =
        T_{m+1}^{(2)}(u) T_{m-1}^{(2)}(u)+
        T_{-2m}^{(1)}(u) .  \label{eq-c2-2}
\end{eqnarray}
Now we briefly mention special cases of 
the results \cite{T1,T2,T3} 
concerning $sl(1|2)$ and point out relation to  
 $C(2)$ case.  
The distinguished simple root system of $sl(1|2)$ 
is \{$\epsilon^{*} -\delta_{1}^{*}, 
\delta_{1}^{*} -\delta_{2}^{*} $\}
(see Figure \ref{dynkin-c2}), 
where $(\epsilon^{*}|\epsilon^{*})=1$, 
$(\epsilon^{*}|\delta_{j}^{*})=(\delta_{j}^{*}|\epsilon^{*})=0$, 
$(\delta_{i}^{*}|\delta_{j}^{*})=-\delta_{ij}$,  
$\epsilon^{*}-\delta_{1}^{*}-\delta_{2}^{*}=0$. 
Let $F_{c}^{(1)}(u)$ and $F_{m}^{(2)}(u)$ be the DVFs whose top term
\footnote{
We assume that these top terms are given respectively 
as follows: 
$\frac{Q_{1}(u-\frac{c}{t_{1}})}{Q_{1}(u+\frac{c}{t_{1}})}$, 
$(-1)^{m} \frac{Q_{2}(u-\frac{m}{t_{2}})}{Q_{2}(u+\frac{m}{t_{2}})}$, 
where $t_{1}=1,t_{2}=-1$ (cf. $C(2)$ case: $t_{1}=2,t_{2}=-1$).
}
carry $sl(1|2)$ weights
\footnote{
Kac-Dynkin labels $\{b_{1},b_{2} \}$ 
of these weights are $\{c,0 \}$ 
and $\{0,m \}$ respectively.
}
 $c\epsilon^{*}$ and $-m\delta_{2}^{*}$ 
respectively.  
By using the functions: 
\begin{eqnarray}
&& \framebox{$1$}=\frac{Q_{1}(u-1)}{Q_{1}(u+1)}, \qquad  
 \framebox{$2$}=\frac{Q_{1}(u-1)Q_{2}(u+2)}{Q_{1}(u+1)Q_{2}(u)} ,
 \nonumber \\ 
&& \framebox{$3$}=\frac{Q_{2}(u-2)}{Q_{2}(u)} 
\end{eqnarray}
and 
\begin{eqnarray} 
 && \framebox{$-1$}=\frac{Q_{1}(u)}{Q_{1}(u-2)},  
 \qquad \framebox{$-2$}=\frac{Q_{1}(u)Q_{2}(u-3)}{Q_{1}(u-2)Q_{2}(u-1)} ,
\nonumber \\ 
&& \framebox{$-3$}=\frac{Q_{2}(u+1)}{Q_{2}(u-1)} , 
\end{eqnarray}
they are given as follows:
\begin{eqnarray}
 F_{0}^{(1)}(u) = F_{0}^{(2)}(u) = 1,
\end{eqnarray}
\begin{eqnarray}
 F_{1}^{(1)}(u) &=& 
\begin{array}{|c|} \hline 
    1 \\ \hline
\end{array}
-
\begin{array}{|c|} \hline 
   2 \\ \hline
\end{array}
-
\begin{array}{|c|} \hline 
    3 \\ \hline
\end{array},
\end{eqnarray}
\begin{eqnarray}
 F_{c}^{(1)}(u)&=&\frac{Q_{1}(u-c)}{Q_{1}(u+c-4)}F_{2}^{(1)}(u+c-2) \\
     &=& \frac{Q_{1}(u-c)}{Q_{1}(u+c+2)}F_{-1}^{(1)}(u+c+1) 
 \quad    c\ne 0,1 (c\in {\bf C}),
\end{eqnarray}
where 
\begin{eqnarray}
 F_{2}^{(1)}(u) &=& 
\begin{array}{|c|c|} \hline 
    1 & 1 \\ \hline 
\end{array}
-
\begin{array}{|c|c|} \hline 
    1 & 2 \\ \hline 
\end{array}
-
\begin{array}{|c|c|} \hline 
    1 & 3 \\ \hline 
\end{array}
+
\begin{array}{|c|c|} \hline 
    2 & 3 \\ \hline 
\end{array}, \nonumber \\ 
 F_{-1}^{(1)}(u) &=& 
\begin{array}{|c|c|} \hline 
    -3 & -2 \\ \hline 
\end{array}
-
\begin{array}{|c|c|} \hline 
    -3 & -1 \\ \hline 
\end{array}
-
\begin{array}{|c|c|} \hline 
    -2 & -1 \\ \hline 
\end{array}
+
\begin{array}{|c|c|} \hline 
    -1 & -1 \\ \hline 
\end{array}.
\end{eqnarray}
Here the spectral parameter $u$ is shifted as 
$u-1, u+1$ from the left to the right.
\begin{eqnarray}
 F_{m}^{(2)}(u) &=& 
 \sum_{\{i_{k}\}}
 (-1)^{\sum_{k=1}^{m}p(i_{k})}
\begin{array}{|c|} \hline 
    i_{1} \\ \hline 
    i_{1} \\ \hline
    \vdots \\ \hline
    i_{m} \\ \hline
\end{array}
\qquad  {\rm for } \quad m \in {\bf Z}_{\ge 1},
\end{eqnarray}
where the spectral parameter $u$ is shifted as 
$u+m-1,u+m-3,\dots, u-m+1$ 
from the top to the bottom; $p(-1)=0,p(-2)=p(-3)=1$. 
Summation is taken over the tableaux 
$\{ i_{j} \}$ ($i_{j}\in \{-1,-2,-3 \}$) 
with the condition: 
 $i_{j} \preceq i_{j+1}$, and 
 $i_{k} \prec i_{k+1}$ if $i_{k+1}=-1$, 
 where we assume $-3 \prec -2 \prec -1$.
Due to the isomorphism $C(2) \simeq sl(1|2)$, 
the following relation will hold: 
\begin{eqnarray}
 T_{m}^{(1)}(u)=F_{\frac{m}{2}}^{(1)}(u), \qquad 
 T_{m}^{(2)}(u)=F_{m}^{(2)}(u). \label{c2-sl12}
\end{eqnarray}
Noting this relation, we can verify 
the functional relation (\ref{eq-c2-1}), i.e. 
\begin{eqnarray}
\hspace{-45pt} &&  
F_{-\frac{m}{2}}^{(1)}(u-\frac{1}{2})
F_{-\frac{m}{2}}^{(1)}(u+\frac{1}{2})  = 
        \left\{
          \begin{array}{ll}
            F_{-1}^{(1)}(u)(F_{1}^{(2)}(u)+1)
             & {\rm for} \quad m=1, \\ 
            F_{\frac{-m+1}{2}}^{(1)}(u) F_{\frac{-m-1}{2}}^{(1)}(u) 
              & {\rm for} \quad m\in {\bf Z}_{\ge 2}.
          \end{array}
        \right. 
\end{eqnarray}
The DVF labelled by the  
(dotted) Young superdiagram
\footnote{(Dotted) Young superdiagram $\mu=(\mu_{1},\mu_{2},\dots)$
 is related to 
Kac-Dynkin label $(b_{1},b_{2})$ of $sl(1|2)$ as follows:
$b_{1}= -\xi_{1}-\mu_{2}^{\prime}, 
b_{2}=\mu_{1}^{\prime}-\mu_{2}^{\prime}$, 
where $\xi_{1}=Max(\mu_{1}-2,0)$.
}
 with shape $(m^{a})$ is given as follows:
\begin{eqnarray}
{\cal F}_{m}^{a}(u)=
{\rm det}_{1\le i,j \le a}({\cal F}_{m+i-j}^{1}(u+a-i-j+1)).
\end{eqnarray}
This satisfies the following functional relation: 
\begin{eqnarray}
{\cal F}_{m}^{a}(u-1) {\cal F}_{m}^{a}(u+1)  =  
{\cal F}_{m-1}^{a}(u) {\cal F}_{m+1}^{a}(u)
+{\cal F}_{m}^{a-1}(u) {\cal F}_{m}^{a+1}(u).
\label{hiro}
\end{eqnarray}
Noting the following relations (\ref{vani}) and (\ref{alta}),
\begin{eqnarray}
{\cal F}_{m}^{a}(u)=0 \qquad {\rm if} \quad 
m \in {\bf Z}_{\ge 3} \quad {\rm and} \quad 
 a \in {\bf Z}_{\ge 2} ,
\label{vani}
\end{eqnarray}
\begin{eqnarray}
 {\cal F}_{2}^{a}(u)= {\cal F}_{a+1}^{1}(u) \qquad 
{\rm for} \quad a \in {\bf Z}_{\ge 1},
\label{alta}
\end{eqnarray}
 we find (\ref{hiro}) reduces to
 the following set of functional relations: 
\begin{eqnarray}
&& {\cal F}_{1}^{a}(u-1) {\cal F}_{1}^{a}(u+1)  =  
{\cal F}_{2}^{a}(u)
+{\cal F}_{1}^{a-1}(u) {\cal F}_{1}^{a+1}(u) 
\quad a \in {\bf Z}_{\ge 1}, \\ 
&& 
 {\cal F}_{2}^{1}(u-1) {\cal F}_{2}^{1}(u+1)  =  
{\cal F}_{1}^{1}(u) {\cal F}_{3}^{1}(u)
+ {\cal F}_{3}^{1}(u), \\ 
&& {\cal F}_{m}^{1}(u-1) {\cal F}_{m}^{1}(u+1)  =  
 {\cal F}_{m-1}^{1}(u) {\cal F}_{m+1}^{1}(u)
\quad m \in {\bf Z}_{\ge 3}.
\end{eqnarray}
This can be rewritten as 
\begin{eqnarray}
\hspace{-45pt} &&  
F_{-m}^{(1)}(u-1) F_{-m}^{(1)}(u+1)  = 
        \left\{
          \begin{array}{ll}
            F_{-2}^{(1)}(u)(F_{1}^{(2)}(u)+1)
             & {\rm for} \quad m=1, \\ 
            F_{-m+1}^{(1)}(u) F_{-m-1}^{(1)}(u) 
              & {\rm for} \quad m\in {\bf Z}_{\ge 2},
          \end{array}
        \right.   \label{dotfun1} \\  
\hspace{-45pt} &&   
 F_{m}^{(2)}(u-1) F_{m}^{(2)}(u+1)  =
        F_{m+1}^{(2)}(u) F_{m-1}^{(2)}(u)+
        F_{-m}^{(1)}(u) \quad  m\in {\bf Z}_{\ge 1},
         \label{dotfun2}
\end{eqnarray}
where 
$F_{-m}^{(1)}(u)={\cal F}_{2}^{m}(u)={\cal F}_{m+1}^{1}(u)$ 
and $F_{m}^{(2)}(u)={\cal F}_{1}^{m}(u)$. 
Noting the relations (\ref{c2-sl12}), we find (\ref{dotfun2}) is 
equivalent to (\ref{eq-c2-2}).
              
\end{document}